\documentstyle[12pt]{article}

\renewcommand{\baselinestretch}{1.5}

\def\({\c c}
\def\|{\'\i}

\def\nl {\par \noindent }

\begin{document}

\begin{center}
{\baselineskip=18pt{\large \bf LIGHT-FRONT QUANTIZED FIELD THEORY: \\ 
{\it  SOME NEW RESULTS} }}
\baselineskip=14pt
\footnote{\baselineskip=12pt 
  Invited talk given at the IX Brazilian School of Cosmology and 
Gravitation, CBPF, Rio de Janeiro, July 1998. To be published 
in the Proceedings, Ed. M. Novello.} 

\vspace{0.9cm}

{\bf Prem P. Srivastava}
\footnote{\baselineskip=12pt {\sl Present Address}:
  {\it Theoretical Physics Group, SLAC, Stanford University, PO Box 4349,
 Stanford, USA.}  \\   
{\sl E-mail}:  prem@slac.stanford.edu,\quad 
prem@lafexsu1.lafex.cbpf.br }\\
\vspace{0.4cm}
{\it  Instituto de F\|sica, UERJ, Rio de Janeiro, RJ, Brasil} \\
%{\it Rua S\~ao Francisco Xavier 524,  
%Rio de Janeiro, RJ, 20550-013, Brasil} \\
and \\
{\it Centro Brasileiro de Pesquisas F\|sicas, Rio de Janeiro, RJ, Brasil}
\end{center}
%\vspace{0.2cm}
\baselineskip=14pt 
\begin{center}
{\bf Abstract}
\end{center}
%\vspace{0.1cm}
%\begin{abstract}
A review is made on some recent studies which support the point of 
view that the 
relativistic field theory quantized on the  light-front (LF), as proposed by 
Dirac, seems to be  more transparent compared to the
 conventional equal-time quantized one. 
 Some ideas following from these studies may be of some relevance 
in the context of the quantization of gravitation theory. 
%for the quantization of the gravitation theory.
%in the curved space-time as well. 

It is argued on general grounds that the LF quantization 
is {\it equally appropriate} as the 
conventional equal-time one and that the two should lead, 
assuming the microcausality principle, to the same 
physical content. This is shown to be true by considering several model 
field theories. The description on the LF 
of the spontaneous symmetry breaking (SSB), (tree level) Higgs mechanism, 
of the {\it condensate} or $\theta$-vacua in the Schwinger model (SM), 
of the absence of such 
vacua in the Chiral SM (CSM), and  of the BRS-BFT quantization of the 
{\it front form} CSM are among the topics discussed. 

The LF phase space is strongly constrained and is different from the 
one in the conventional theory. The removal of the constraints by 
following the Dirac 
procedure results in  a substantially 
reduced number of independent operators. The discussion of the 
physical Hilbert space and the vacuum becomes more tractable.

Some comments on the irrelevance, in the quantized 
field theory, of the fact that the hyperplanes $x^{\pm}=0$ 
constitute characteristic surfaces of the hyperbolic partial 
differential equation are also made. The LF theory quantized on, say, the 
$x^{+}=const.$ hyperplanes seems to contain in it the information 
on the equal-$x^{-}$ commutators as well. 

A theoretical reaffirmation  of 
the universally  accepted notion 
that the experimental data is to be confronted with the
predictions of a classical theory model only after it has been 
upgraded through its quantization seems to emerge. 
The LF quantization promises to be a 
powerful tool, complementary to the functional integral  method, 
 for handling the nonperturbative calculations.

\newpage

\nl {\bf 1- Introduction} \\
%\section{Introduction}\label{intro}

\medskip
 
Dirac \cite{dir}, in his paper, in 1949, discussed the unification, 
in a relativistic theory,  of 
the principles of the quantization and the special relativity theory. 
The Light-Front (LF) 
quantization which studies the  relativistic 
quantum dynamics of  physical system on the hyperplanes  
: $x^{0}+x^{3}\equiv {\sqrt{2}}x^{+}=const.$,  called 
the {\it front form} theory, was also proposed there. 
The {\it instant form} or the 
conventional equal-time theory 
on the contrary uses the $x^{0}=const.$ hyperplanes.  
The LF coordinates $x^{\mu}: (x^{+},x^{-},x^{\perp} 
)$,  where $x^{\pm}=(x^{0}{\pm} x^{3}) 
/{\sqrt 2}=x_{\mp}$ and   $ x^{\perp} = 
(x^{1}, x^{2})$,   are convenient to use in the {\it front form}
theory.  
They are  {\it not related by a Lorentz transformation} 
to the coordinates $(x^{0}\equiv t,x^{1},x^{2},x^{3})$ 
usually employed in the {\it instant form } theory and as such the 
descriptions of the same physical content of a dynamical theory 
on the LF may come out to be different from that given in the conventional 
treatment. The LF quantized field theory may hence be of some 
relevance in the understanding of the unification of the 
principles of the quantization with that of the general 
covariance\footnote{\baselineskip=12pt 
We recall the Kruskal-Szekers coordinates which 
threw a new light on the problem of the Schwarzshild singularity.}.

We will make the {\it convention} to regard\footnote{\baselineskip=12pt 
The coordinates $x^{+}$ and $x^{-}$ appear in a symmetric fashion 
and we note that $\left[x^{+},{1\over i}\partial^{-} \right]
= \left[x^{-},{1\over i}\partial^{+} \right]=i $ where $\partial^{\pm}=
\partial_{\mp}=(\partial^{0}\pm \partial^{3})/\sqrt {2}$ etc.. }
$x^{+}\equiv \tau$ as the 
LF-time coordinate while $x^{-}\equiv x$ as the {\sl longitudinal 
spatial} coordinate. The temporal  evolution in $x^{0}$ or 
$x^{+}$ of the system is 
generated by Hamiltonians which are different 
in the two {\it forms} of the theory. 
The LF components, with $\mu=+,-,1,2$,  of any 
tensor are defined likewise.

Consider \cite{pre} the invariant distance between two spacetime points 
: $ (x-y)^{2}=(x^{0}-y^{0})^{2}-(\vec x-\vec y)^2= 2 
(x^{+}-y^{+}) (x^{-}-y^{-}) - (x^{\perp}-y^{\perp})^{2}$. 
On an equal $x^{0}=y^{0}=const. $ hyperplane the points have  
spacelike separation  except for if  they 
are {\it coincident} when it becomes lightlike one.  
On the LF with $x^{+}=y^{+}=const.$ 
the distance becomes  {\it independent of}  $(x^{-}-y^{-})$ and 
the seperation is again spacelike; it becomes lightlike one 
when  $x^{\perp}=y^{\perp}$ but with the difference that 
now the points need {\it not}  
necessarily be coincident along the longitudinal direction. 
The LF field theory hence need not necessarily be local 
in $x^{-}$, even if the corresponding 
{\it instant form} theory is formulated as a local one. 
For example, the commutator 
$[A(x^{+},x^{-},{x^{\perp}}),B(0,0,0^{\perp})]_{x^{+}=0}$ 
of two scalar observables would vanish on the grounds of
microcausality principle if   
$ x^{\perp}\ne 
0$ when  $x^{2}\vert_{x^{+}=0}$ is spacelike. 
Its value  would  hence be proportional to  $\,\delta^{2}(x^{\perp})\, $ 
and a finite number of its derivatives,  
implying locality only in $x^{\perp}$ but not necessarily so 
in $x^{-}$. Similar arguments in 
the {\it instant form} theory lead to the locality 
in all the three spatial coordinates. 
In view of the microcausality both  of the commutators 
 $[A(x),B(0)]_{x^{+}=0}$ and 
$[A(x),B(0)]_{x^{0}=0}$ are nonvanishing    
only on the light-cone. 

We remark that in the LF 
quantization we time order with 
respect to   $\tau$ rather than  $t$.  
The  microcausality principle, 
however, ensures that the retarded commutators  
$[A(x),B(0)]\theta(x^{0})$ and  $[A(x),B(0)]\theta(x^{+})$,  
which appear \cite{ryd} in the S-matrix elements of relativistic 
field theory, do not  lead to disagreements in the two formulations. 
In the regions 
$x^{0}>0, x^{+}<0$ and $x^{0}<0, x^{+}>0$, where the commutators  
seem different  the $x^{2}$  is spacelike and both of them vanish. 
Hence, admitting the {\it microcausality
principle} to hold,  the LF hyperplane  
seems  {\it equally appropriate} as the conventional  one 
of the {\it instant form} theory for the canonical quantization.

The structure of the {\it LF phase space}, however, 
 is different from that of the one in the  conventional theory. 
Consequently, we may require on the LF a different description of 
to the same physical content as found in the conventional treatment. 
For example, the SSB needs a different description \cite{pre} 
or mechanism on the LF when compared with the conventional one. 
The broken continuous symmetry is now inferred 
from the study of the residual unbroken symmetry of the  LF Hamiltonian 
operator while the symmetry of the LF vacuum remains intact. The 
expression 
which counts the number of Goldstone bosons 
present in the theory, a physical content, comes out to be the same as 
that found in the 
equal-time quantized theory. A new proof of the
Coleman's theorem \cite{col} on the absence of the Goldstone bosons 
in two dimensional
theory also emerges \cite{pre}  easily on the LF. 
The LF vacuum 
is generally  found to be simpler \cite{bro, ken} and in many cases 
the interacting theory vacuum 
is seen to coincide\footnote{\baselineskip=12pt 
In fact, in many cases the interacting theory vacuum 
may coincide with the perturbation theory one. 
This results from the fact that  momentum 
four-vector  is now given by $(k^{-},k^{+},k^{\perp})$ where 
$k^{\pm}=(k^{0}{\pm}k^{3})/{\sqrt 2}$.  Here 
$k^{-}$ is the LF energy while $k^{\perp}$ and  $k^{+}$  indicate 
the transverse and the {\it longitudinal} components of the  momentum respectively.  
For a free massive particle on the mass shell we have  $2 k^{-}k^{+}=
({k^{\perp}}^{2}+m^{2}) > 0$ so that  
  $k^{\pm}$ are both positive definite when $ k^{0}>0$. 
The conservation of the total 
longitudinal momentum does not permit the excitation of massive 
quanta by the LF vacuum. We  require $k^{+}\to 0$ for each particle (and
antiparticle) entering the ground state, which has vanishing total momentum. 
Such configurations constitute a point with zero measure in the LF phase
space and may \cite{bro} not be of relevance in many cases. However, it should
be noted that when dealing with the momentum space integrals, say,
the loop integrals, a significant contribution may arise precisely
from such a corresponding configuration in the integrand; the reason
being that we have to deal with the products of several
distributions. On the other hand, $(k^{1},k^{2},k^{3})$ may take positive
or negative values and we may construct in the conventional theory 
eigenstates of zero momentum  with an arbitrary number of particles 
(and antiparticles) which may mix with the vacuum state, with no 
particles, to form the ground state.}.

An important advantge pointed out by Dirac of {\it front form} 
theory is that here {\it seven} out of the ten   
Poincar\'e generators are {\it kinematical}, e.g., they leave 
the plane $x^{+}=0 $ invariant \cite{dir}. They are 
$P^{+},P^{1},P^{2},\, 
M^{12}=-J_{3},\, M^{+-}= M^{03}= -K_{3},\,
M^{1+}=(K_{1}+J_{2})/\sqrt{2}$ and 
$ M^{+2}=(K_{2}-J_{1})/\sqrt{2}$. 
In the conventional theory  
 only six such ones, {\it viz.}, ${\vec P}$ and  
$M^{ij}=-M^{ij} $,  leave the hyperplane $x^{0}=0$ 
invariant. 
In fact, in the standard notation $K_{i}=-M^{0i}, J_{i}=-(1/2)\epsilon_{ijk}
M^{kl}, i,j,k=1,2,3$ and the generator 
$K_{3}$ is dynamical one in the {\it instant form} theory. It  
is in contrast kinematical in the {\it front form} theory where it 
generates  the scale 
transformations of the LF components of $x^{\mu}$, 
$P^{\mu}$ and  $M^{\mu\nu}$, with  $\mu,\nu=+,-,1,2$.  
It is also worth remarking that the $\,+\,$ component 
of the Pauli-Lubanski pseudo-vector $W^{\mu}$ is special in that it 
contains only the LF kinematical generators. 
This suggests us to define the  {\it LF Spin operator} 
by  ${\cal J}_{3}=-W^{+}/P^{+}$. The other two components of 
$\vec {\cal J}$ are shown to be 
 ${\cal J}_{a}=-({\cal J}_{3}P^{a}
+W^{a})/\sqrt{P^{\mu}P_{\mu}}, \,  a=1,2$, which, however,  do  
carry\footnote{\baselineskip=12pt 
 See refs. \cite{pre} and \cite{bro}.} in them 
 also the  LF {\it dynamical generators} 
$P^{-}, M^{1-}, M^{2-} $.

Another distinguishing feature of the {\it front
form}  theory is that it gives rise generally to 
a (strongly) constrained dynamical system \cite{dir1} which  leads 
to an appreciable  reduction in the number of independent operators 
on the phase space. The 
vacuum structure, for example,  then becomes more tractable and the 
computation of  physical quantities  simpler.  This is 
verified in the recent study of the LF quantized SM  
\cite{pre1} and the Chiral SM (CSM) discussed below where  we are led 
directly to  the {\it physical Hilbert space}, once the
constraints are taken into account by following the Dirac procedure 
\cite{dir1}. 

We recall that the 
LF field theory was rediscovered in 1966 by Weinberg \cite{wei} in his 
Feynman rules adapted for infinite momentum frame. It  was 
demonstrated \cite{kog} latter that these  rules, in fact, correspond to 
the {\it front form} quantized theory.  
It was also employed successfully in the {\it nonabelian bosonization
} of the field theory of N free Majorana fermions,  
where the corresponding LF current algebra  was compared \cite{wit} 
with the one in the bosonized theory described by the WZNW action 
at the critical point. 
 The interest in LF quantization  
has been revived \cite{bro, ken} also  
due to the difficulties encountered 
in the computation, in the conventional 
framework, 
 of the nonperturbative effects in the context of  QCD 
and  the problem of the relativistic bound states of light 
fermions \cite{ken, bro} 
in the presence of the complicated vacuum. 
Studies show  that the application of 
Light-front Tamm-Dancoff method  may be feasible here.
 The technique of the 
regularization on the lattice has been 
quite successful for some problems but it cannot handle, for example, the 
light ( chiral) fermions and  has not  been able yet 
to demonstrate, for example,  the confinenment of 
quarks. 
The problem of reconciling 
the standard constituent quark model and the QCD to 
describe the hadrons is also not satisfactorily resolved. In the former 
we employ few valence quarks while in the latter the QCD vacuum state itself 
contains, in the conventional theory, 
an infinite sea  of constituent quarks and gluons ( partons) 
with the density 
of low momentum constituents getting very large in view of the infrared 
slavery. 
  The {\it front form} dynamics may serve as a 
complementary tool to study such probelms  since  we have a simple vacuum here while the 
complexity of the problem is now transferred to the 
LF Hamiltonian. In the case of the scalar field theory, for example, 
the corresponding LF  Hamiltonian is, in fact, 
found \cite{pre}  
to be nonlocal due to the appearence of 
{\it constraint equations} on the {\it LF phase space}. 

We discuss here only some of the interesting conclusions 
reached from the detailed  study of some model relativistic theories on 
the LF and 
where the standard Dirac procedure for constrained dynamical systems 
is followed in order to build the self-consistent Hamiltonian 
formulation. Among some of the results obtained  we find that

\begin{itemize}
\item 
The LF hyperplane is {\it equally appropriate} as the conventional equal-time 
one for the field theory quantization. 
\item 
The  
hyperplanes $x^{\pm}=0$ define the characteristic surfaces of a 
hyperbolic partial differential equation. From the mathematical theory of 
classical partial differential equations \cite{sne}  
it is known that the Cauchy initial value problem 
would require us to specify the  data on both the hyperplanes. 
From our studies 
we conclude that 
it is sufficient in the {\it front form} theory  to choose, 
 as proposed by Dirac \cite{dir},  one of the two LF hyperplanes 
for canonically quantizing the theory. 

- In the quantized theory the equal-$\tau$  commutators of 
the field operators,   at a fixed 
initial LF-time, form now  a part of the initial data instead and we 
deal with operator differential equations. 

- The studies show that the information on the commutators on 
the other characteristic 
hyperplane are already contained \cite{pre1} 
in the quantized theory and need not thus be specified separately. 

\item The inherent symmetry with regard to equal-$x^{\pm}$ 
commtators along with the reduced number of independent 
field operators which survive after the constraints are taken care of 
seem to be responsible for the  very transparent discussion on the LF.

\item The physical content following from the {\it front form} theory 
is the same, even though 
arrived at through different description on the LF, when compared with the 
one in the {\it instant form} case.

\item In the conventional treatment we sometimes are required to 
introduce external 
constraints in the theory based on physical considerations, say, 
when describing the spontaneous symmetry breaking. Many of the analogous 
constraints may be shown to be already icorporated  in 
the quantized theory considered on the LF.

\item A theoretical demonstration of the well accepted notion 
that a  classical model field 
theory must be upgraded first through its quantization 
before we confront it  with the 
experimental data, seems to emerge. 

\item The recently proposed BRS-BFT \cite{bat} quantization procedure is 
extended straightforwardly 
on the LF as well as is illustrated below 
 in the context of CSM (Appendix D).

\item Topological considerations often employed in the context of the 
functional  
integral techniques, where the Euclidean theory action is ususally 
employed, 
also seem to have their counterpart, though now interpreted 
differently.  This is suggested, for example, from the studies of the 
LF quantized SM and CSM.  

\end{itemize}

For illustration purposes we discuss in the following Sections the description 
on the LF of the spontaneous symmetry breaking (SSB) 
and of the structure of the vacuum state in the 
CSM while some other topics related to the {\it front form} theory are 
collected in the Appendices.

\bigskip

\nl {\bf 2-  Spontaneous Symmetry Breaking Mechanism on the LF} \\
%\section{Spontaneous Symmetry Breaking Mechanism on the LF}
% SSB Mechanism. Continuum limit of discretized LF 
%quantized theory. Nonlocality of LF Hamiltonian }

On  the canonical  quantization of   the {\it instant form} scalar 
field Lagrangian theory we obtain as well known 
the Hamiltonian and the commutation relations among the 
field operators. The description of, say, the tree level SSB emerges when we 
require also (e.g., introduce {\sl external} constraints ), based on physical 
considerations, that the $\phi_{classical}\equiv \omega$ corresponds 
to the minimum of the Hamiltonian functional. The {\it front form} of the same 
theory describes \cite{pre} a constrained dynamical system and the canonical 
Hamiltonian framework, which may be quantized by the correspondence 
principle, is shown to contain in it a new ingredient in the form of 
the {\it constraint equations}, 
in addition to the  Hamiltonian  and commmutators among the field operators.  
The constraint equations  may also be derived 
from the Lagrange equations of motion
\footnote{\baselineskip=12pt  That the   constraint equations can  
 be derived simply by integrating the Lagrange equations of motion  
over the longitudinal 
spatial coordinate $x^{-}$ was 
noted also in: P.P. Srivastava, {\it On spontaneous symmetry breaking 
mechanism on the light-front quantized field theory}, Ohio-State preprint 
91-0481, Slac database PPF-9148 (see also 92-0012, PPF-9202), November '91; 
available as {\it scanned} copies. 
In fact, Dirac in his paper does consider some examples where the constraints 
on the form of the potential are required if we would like to   
unify in the dynamical theory 
the principles of the quantization and the relativistic invariance. 
It is interesting to note that soon after in 1950-52 he formulated also the 
systematic method (Dirac procedure) for constructing Hamiltonian formulation 
for  a constrained dynamical system.}.  
The new ingredient permits us to describe \cite {pre} SSB on the LF 
without requiring us to deal with the difficult task of introducing 
constraints on the LF based on physical considerations. Some of them 
may be shown to be already incorporated \cite{pre} 
 in the formulation itself due to the requirement of 
the selfconsistency \cite{dir1}.

The existence of the  continuum limit of the 
Discretized Light Cone Quantized 
(DLCQ) \cite{pauli} theory, the nonlocal nature of the LF Hamiltonian, and the 
description of the SSB on the LF were 
clarified \cite{pre, pre2} only recently. 

 Consider first the two  
dimensional case of single real scalar theory  with the Lagrangian 
${\cal L}= \; \lbrack
{\dot\phi}{\phi^\prime}-V(\phi)\rbrack$. 
 Here $\tau\equiv x^{+}=(x^0+x^1)/{\sqrt2}$,
  $x\equiv x^{-}=(x^0-x^1)/{\sqrt2}$, $\partial_{\tau}\phi=\dot\phi , 
       \partial_{x}\phi={\phi}'$, and $d^2x=d\tau dx$. It is the simplest 
example of a constrained field theory.  
The eq. of motion, 
$\,\dot{\phi^\prime}=(-1/2)\delta V(\phi)/\delta \phi $, shows that 
$\phi=const. $ is a possible solution. 
We propose  to make the following 
{\it separation}\footnote{\baselineskip=12pt  
It was first proposed in the ref. cited in the previous footnote; 
in $3+1 $ dimensions the separation is: 
$\;\phi(\tau,x^{-},x^{\perp})=
\omega(\tau,x^{\perp})+\varphi(\tau,x^{-},x^{\perp})$. See papers contribuited to {\it XXVI Intl. Conference on High energy Physics, Dallas, Texas}, August 
'92, {\it AIP Conf. Proc.}, 272 (1993) 2125, Ed. J.R. Sanford.}  
$\;\phi(\tau,x)=\omega(\tau)+\varphi(\tau,x)$ where 
the $\omega(\tau)$ is the {\it dynamical variable}  
representing the {\it bosonic condensate} and $\varphi(\tau,x)$ describes 
(quantum) {\it fluctuations} above it. We set $\int dx^{-} \varphi(\tau,x)=0$ 
so that the fluctuation field carries no zero momentum mode in it. 
Subsequently, we apply the standard Dirac procedure in order to construct a 
selfconsistent Hamiltonian formulation which may be quantized canonically.

We are led \cite{pre} to 

\begin{eqnarray}
\left[\varphi(x,\tau),\varphi(y,\tau)\right]&=&-{i\over 4}\epsilon(x-y),\\
\nonumber \\
\left[\omega(\tau),\varphi(x,\tau)\right] &=&0,
\end{eqnarray}
and 
\begin{equation}
{H^{lf}\equiv 
 P^{-}}= \int  dx \,\Bigl [\omega(\lambda\omega^2-m^2)\varphi+
{1\over 2}(3\lambda\omega^2-m^2)\varphi^2+
\lambda\omega\varphi^3+{\lambda\over 4}\varphi^4 
\Bigr], 
\end{equation}
along with the (second class) constraint equation
\begin{eqnarray}
lim_{R\to\infty} 
{1\over R}\int_{-R/2}^{R/2} dx \,V'(\phi)
& \equiv &  \omega(\lambda\omega^2-m^2)+ lim_{R\to\infty} 
\nonumber \\
&  &{1\over R}
 \int_{-R/2}^{R/2} dx \Bigl[ \,(3\lambda\omega^2-m^2)\varphi + 
 \lambda (3\omega\varphi^2+\varphi^3 ) \,
\Bigr]=0 
\end{eqnarray}
where we have assumed $V(\phi)=
(\lambda/4)(\phi^{2}-m^{2}/\lambda)^{2}$, $\lambda \geq 0$, $m\neq 0$.

Eliminating $\omega$ would lead to a nonlinear and nonlocal Hamiltonian 
in the {\it front form} theory even when the scalar theory is written above 
a local one in the conventional {\it instant form} formulation.

At the tree or classical level $\varphi$ are bounded ordinary functions 
in $x^{-}$ and only the first term survives in the 
constraint equation leading to 
 $V'(\omega)=0$, which is the same as found in the conventional theory. 
There it is essentially added to the theory, 
on the physical considerations which require the energy functional to 
attain is minimum (extremam) value. 
The stability property, say,  of a particular constant solution 
may be inferred as usual from the  
classical partial differential equation of motion. For example,  
$\omega=0$ is shown to be an unstable solution 
for the potential $V$ considered above 
while the other two 
solutions with $\omega\neq 0$ give rise to the  stable 
phases\footnote{\baselineskip=12pt A similar analysis of the corresponding 
{\it partial} differential  equations 
 in the conventional treatmnet can also 
be made; the Fourier transform theory is convenient to use.}.

The  construction of the Hamiltonian formulation using the 
Dirac method \cite {dir1} is a straightforward 
exercise. 
We may use \cite {pre} the continuum 
formulation directly or proceed from the DLCQ \cite {pauli}, 
and take  its infinitie volume 
limit \cite{pre2} to obtain the same results. The canonical quantization 
is performed  via the  
correspondence which relates the final Dirac brackets with the 
commutators (or anticommutators). We note that the Dirac procedure 
when applied to the scalar theory written  in the continuum 
 shows that the variable $\omega$ is  a c-number or a 
background field; in the theory described in 
finite volume, however,  its commutator with $\varphi$ is 
nonvanishing \cite{pre} and as such it is a q-number  operator. We stress 
that in our discussion the condensate variable is introduced as a dynamical 
variable and we let the Dirac procedure decide if it comes out as a c- or 
q-number. In the SM it comes out to be an operator rather than a background field.

In the quantized theory the constraint equation above shows that the value of  
$\omega$ would be altered from its tree level value 
due to the quantum corrections arising from the other terms. It is, 
in fact,  straightforward to renormalize the theory, say,  
up to one-loop order 
by employing the 
Dyson-Wick  expansion. We do not need to solve the constraint first which would give rise to a very complicated LF Hamiltonian. It is more convenient to 
derive \cite{pre} the renormalized constraint eqn. which together with the 
expression of mass renormalization  condition give us two eqns. which may be 
used  to study \cite{pre} the phase transition as conjectured by Simon and 
Griffiths \cite{simon}. 

In view of the LF commutator above the scalar field has the 
LF momentum space expansion :\quad 
$\varphi(x,\tau)= {(1/{\sqrt{2\pi}})}\int dk\; 
{\theta(k)}\;
[a(k,\tau)$ $e^{-ikx}+{a^{\dag}}(k,\tau)e^{ikx}]/(\sqrt {2k})$, 
were $a(k,\tau)$ and ${a^{\dag}}(k,\tau)$ 
satisfy the canonical equal-$\tau$ commutation relations, 
$[a(k,\tau),{a(k^\prime,\tau)}^{\dag}]=\delta(k-k^\prime)$ etc.. 
The vacuum state is defined 
by  $\,a(k,\tau){\vert vac\rangle}=0\,$, 
$k> 0$ and the tree level  
description of the {\it SSB} is  given as
follows. The values of $\omega=
\,{\langle\vert \phi\vert\rangle}_{vac}\;$ obtained from  
$V'(\omega)=0 $ characterize 
the different   vacua in the theory. 
Distinct  Fock spaces corresponding to different values of $\omega$ 
are built as usual by applying the creation operators on the corresponding
vacuum state.  The $\omega=0$ corresponds to a {\it symmetric phase} 
since the Hamiltonian operator is then symmetric under 
$\varphi\to -\varphi$. For $\omega\ne0$ this symmetry is 
violated and the  system is 
in a {\it broken or asymmetric phase}.  

The extension  to 
$3+1$ dimensions and to the global continuous symmetry 
is straightforward\footnote{\baselineskip=12pt  See 
Nuovo Cimento A107 (1994) 549 and 
ref. \cite{pre}.}. 
Consider  real scalar fields 
$\phi_{a} (a=1,2,..N)\,$ which form  an isovector of global 
internal symmetry group 
 $O(N)$. We now  write\footnote{\baselineskip=12pt 
In general $\phi_{a}(x^{-},x^{\perp},\tau)
=\omega_{a}(x^{\perp},\tau)+\varphi_{a}(x^{-},x^{\perp},\tau)$ and the 
$x^{\perp}$ 
dependent tree level configurations (e.g. kinks etc.)  
are determined from   
$\left[V'_{a}(\omega)-\partial^{\perp}.\partial^{\perp}\omega_{a}\right]=0$.}
 $\phi_{a}(x,x^{\perp},\tau)
=\omega_{a}+\varphi_{a}(x,x^{\perp},\tau)$ and 
the Lagrangian density is  ${\cal L}=[{\dot\varphi_{a}}{\varphi'_{a}}-
{(1/ 2)}(\partial_i\varphi_{a})(\partial_i\varphi_{a})-V(\phi)]$,  
where $i=1,2$ indicate the transverse space directions. The Taylor series 
expansion of the constraint equations $\beta_{a}=0$ gives a set of coupled
equations $R\,V'_a(\omega)+
\,V''_{ab}(\omega)\int dx \varphi_{b}+\,
V'''_{abc}(\omega)\int dx \varphi_b\varphi_c/2+...=0$. Its discussion at 
the tree level leads to the conventional theory results. 
The  LF symmetry generators are found to be 
$G_{\alpha}(\tau)=
-i\int d^{2} {x^{\perp}} dx \varphi'_{c}(t_{\alpha})_{cd}\varphi_{d}$ 
$=\,\int d^{2}{k^{\perp}}\,dk \, \theta(k)\, 
{a_{c}(k,{k^{\perp}})
^{\dag}} (t_{\alpha})_{cd} a_{d}(k,{k^{\perp}})$ where 
$\alpha,\beta=1,2,..,N(N-1)/2\, $,  are the group indices, 
$t_{\alpha}$ are hermitian and antisymmetric generators of $O(N)$, and 
${a_{c}(k,{k^{\perp}})^{\dag}}$ ($ a_{c}(k,{k^{\perp}})$) is creation (
destruction)  operator  contained in the momentum space expansion 
of $\varphi_{c}$. These  are to be contrasted with the generators 
in the equal-time theory,  
$ Q_{\alpha}(x^{0})=\int d^{3}x \, J^{0}
=-i\int d^{3}x (\partial_{0}\varphi_{a})(t_{\alpha})_{ab}\varphi_{b} 
-i(t_{\alpha}\omega)_{a}\int d^{3}x 
({{d\varphi_{a}}/ dx_{0}})$.  
All  the symmetry generators thus  
annihilate the LF vacuum  and the SSB is  
now seen  in the 
broken symmetry of the quantized theory Hamiltonian. The criteria for 
the counting of the number of Goldstone bosons on the LF is found 
to be the same as in the conventional theory. 
In contrast, the first term 
on the right hand side of $Q_{\alpha}(x^{0})$ does  annihilate  
the conventional theory vacuum  but 
the second term  gives now  non-vanishing contributions  
for some of the (broken) generators. The  symmetry of the
conventional theory vacuum 
is thereby broken  while the quantum Hamiltonian remains invariant.  
The physical content  of SSB in the {\it instant form}
and the {\it front form}, however, is the same 
though achieved  by different descriptions. Alternative proof  
 on the LF,  
in two dimensions, can be given  
of the Coleman's theorem related to the absence of 
Goldstone bosons; we are unable \cite{pre} to implement 
the second class constraints over the phase space. 
We remark that the simplicity of the LF vacuum is in a sense 
compensated by the involved nonlocal Hamiltonian. The latter, however, 
may be treatable using advance computational techniques. Also in connection with renormalization it may not be necessary\footnote{\baselineskip=12pt  See  
Nuovo Cimento A108( 1995) 35.} first to solve all the 
constraint equations.

To summarize, the simple  procedure of separating first the condensate 
variable $\omega$ in the 
scalar field  before applying the Dirac procedure is found to be successful 
also in describing \cite{pre} the phase transition in two dimensional scalar theory, 
the SSB of continuous symmetry, a new proof of the Coleman's theorem and 
the tree level Higgs mechanism. It is again found  successful in showing 
\cite{pre1} 
the 
emergence on the LF of the $\theta$-vacua along with their 
continuum normalization in the bosonized 
SM while explaining at the same time their absence in the CSM. 
The condensate variable, we remind,  is introduced as a dynamical variable and 
we let the Dirac procedure  decide 
if it comes out to be a c-number (background field) or 
a q-number operator in the quantized field theory. It is shown \cite{pre} 
  to be c-number 
in the scalar theory studied in the continuum while it is 
an operator in the SM  whose eigenvalues characterize the $\theta$-vacua. 
In the next Sec. we discuss in some detail the vacuum structure 
in the CSM which illustrates the remarkable transparency attained 
in the discussion on the LF.

It is worth remarking that 
the LF formulation is inherently symmetrical with respect to 
$x^{+}$ and $x^{-}$ 
and it is a matter of convention that we take the plus component as
the LF {\sl time} while the other as a spatial coordinate. 
The  theory quantized at $x^{+}=const.$  hyperplanes 
seems already to incorporate in it the information on the 
equal-$x^{-}$ commutation relations.  We need to quantize the theory, 
as suggested by Dirac,  only on one of the LF hyperplanes. Consider,  
for example, the free scalar theory for which

$$\varphi(x^{+},x^{-})= {1\over{\sqrt{2\pi}}}\int_{k^{+}>0}^{\infty}\,
 {dk^{+}\over{\sqrt{2k^{+}}}}
\Bigl [ a(k^{+}) e^{-i(k^{+}x^{-}+k^{-}x^{+})} +
        a^{\dag}(k^{+}) e^{i(k^{+}x^{-}+k^{-}x^{+})} \Bigr ]$$

\noindent with $\left [a(k^{+}),{a(l^{+})}^{\dag}\right]=\delta(k^{+}-l^{+})$ etc. and
$2k^{+}k^{-}=m^{2}$.  We find easily 

$$\left[\varphi(x^{+},x^{-}),
\varphi(y^{+},x^{-})\right ]={1\over{2\pi}} \int_{k^{+}>0}^{\infty}
{dk^{+}\over {2k^{+}}}\Bigl [e^{ik^{-}(y^{+}-x^{+})}-
e^{-ik^{-}(y^{+}-x^{+})} \Bigr ].$$

\noindent We may change the integration variable to $k^{-}$ by 
making use of $k^{-}dk^{+}+k^{+}dk^{-}=0$ and employ the
integral representation 
$\epsilon(x)= (i/\pi){\cal P}\int_{-\infty}^{\infty} (d\lambda/\lambda) \,
e^{(-i\lambda x)}$ to arrive at the equal-$x^{-}$ commutator 

$$\left[\varphi(x^{+},x^{-}),
\varphi(y^{+},x^{-})\right]= -{i\over 4} \epsilon(x^{+}-y^{+})$$

\noindent The above field expansion on the LF, in 
contrast to the equal-time case, does not 
involve the mass parameter $m$ and 
the same result follows in the massless case also  if we 
assume that $k^{+}=l^{+}$ implies $k^{-}=l^{-}$. 
Defining  the right  and the left movers  by 
$\varphi(0,x^{-})\equiv \varphi^{R}(x^{-})$, 
and $\varphi(x^{+},0)\equiv \varphi^{L}(x^{+})$ we obtain  
$\left [\varphi^{R}(x^{-}),\varphi^{R}(y^{-})\right]=(-i/4)\epsilon(x^{-}-y^{-})$
while $\left[\varphi^{L}(x^{+}),\varphi^{L}(y^{+})\right]=
(-i/4)\epsilon(x^{+}-y^{+})$. The symmetry under discussion is responsible 
for an appreciable simplification found in the recent study of the 
gauge theory SM and the  CSM on the LF discussed in the next Sec..

\bigskip
%\begin{sloppypar}

\nl {\bf 3- Bosonized CSM on the LF. Absence of $\theta$-vacua} \cite{pre5}\\

%\section{Bosonized CSM on the LF. Absence of $\theta$-vacua}\label{bosmod}

The  Lagrangian density of the chiral $QED_{2}$ or CSM  model is

\begin{equation}
{\cal L}= -{1\over 4}F^{\mu\nu}F_{\mu\nu} 
+  {\bar\psi}_{R}\,i\gamma^{\mu}\partial_{\mu}
\psi_{R}+ {\bar\psi}_{L}\,\gamma^{\mu}(i\partial_{\mu}+2e\sqrt{\pi}
 A_{\mu})\psi_{L},
 \end{equation}
where\footnote{\baselineskip=12pt  Here 
$\gamma^{0}=\sigma_{1}$, $ \gamma^{1}=i\sigma_{2}$, 
$\gamma_{5}=-\sigma_{3}$, 
  $x^{\mu}:\,(x^{+}\equiv \tau,x^{-}\equiv x)$ 
with ${\sqrt 2}x^{\pm}={\sqrt 2}x_{\mp}=(x^{0}{\pm} x^{1})$, 
$A^{\pm}=A_{\mp}=(A^{0}\pm A^{1})/
{\sqrt 2}$,  
$\psi_{L,R}=P_{L,R}\;\psi $, $P_{L}=(1-\gamma_{5})/2$, 
$P_{R}=(1+\gamma_{5})/2$, 
$\bar\psi= \psi^{\dag}\gamma^{0}$. 
}.  $\psi= \psi_{R}+\psi_{L}$ is a two-component spinor field and
$A_{\mu}$ is the abelian gauge field, 
$\gamma_{5}\psi_{L}=-\psi_{L}$, and $\gamma_{5}\psi_{R}=\psi_{R}$. 
The classical Lagrangian  is  
 invariant under the local $U(1)$ gauge transformations $A_{\mu}\to 
A_{\mu}+\partial_{\mu}\alpha/(2\sqrt{\pi}e)$, $\psi\to [P_{R}+
e^{i\alpha}P_{L}] \psi $ and under the global 
$U(1)_{5}$ chiral transformations $\psi\to exp(i\gamma_{5}
\alpha)\,\psi $. 

The  model under study can be solved completely using the technique of
bosonization. The latter consists in the replacement of a known
system of fermions with a theory of bosons which has a completely
equivalent physical content, including, for example, identical
spectra, the same current commutation relations and the
energy-momentum tensor when expressed in terms of the currents. The 
bosonized  version is convenient to study the vacuum structure
and it was shown \cite{jac} to be   given by 
\begin{equation} 
    S = \int d^2x \left[ -{{1}\over {4}}F_{\mu \nu}F^{\mu \nu} 
                          +{{1}\over{2}}\partial_{\mu}\phi\partial^{\mu}\phi
                          +eA_{\nu}(\eta^{\mu \nu}
                          -\epsilon^{\mu \nu})\partial_{\mu}\phi
              +{{1}\over{2}}ae^{2}A_{\mu}A^{\mu}\right] 
\end{equation}
Here the explicit mass term for the gauge field 
parametrized by the constant parameter $a$ represents a 
regularization ambiguity 
and the breakdown of $U(1)$ gauge symmetry. The action may be
derived by the functional integral method or by the canonical
quantization.

We make 
  the {\it separation}:     
$\,\phi(\tau,x^{-})= \omega (\tau)+\varphi(
\tau,x^{-})$ and follow it through the application of the 
Dirac method as done with the SSB above. 
On the other hand in the bosonized SM on the LF we recall \cite{pre1} that 
$\omega(\tau)$  turned out 
to be  q-number operator and its  eigenvalues were  shown \cite{pre1} 
to characterize  the {\sl condensate} or $\theta$-vacua \cite{low}, 
which were shown also 
to emerge naturally with a continuum normalization in contrast to what 
found in the conventional equal-time treatment.  
We  remind \cite{pre1} also that  the {\it chiral transformation} 
is defined by:\quad $\omega\to \omega+const., \; \varphi\to \varphi$,  
and $A_{\mu}\to A_{\mu}$. This ensures  
that the {\it boundary conditions} on the 
$\varphi$ are kept   
unaltered under such transformations and our mathematical framework 
may be considered {\it well posed}, before we proceed to build 
the canonical Hamiltonian framework.

Written explicitly 
the action takes the following form on the LF 
\begin{equation} 
 S = \int\, d^2x \;\left[{\dot\varphi}\varphi'+
{1\over 2}({\dot A}_{-}-{A'_{+}})^{2}+ a e^{2}[A_{+}+{2\over {ae}}(
\dot\omega+\dot\varphi)] A_{-} \right] 
\end{equation}
where an overdot (a prime) indicates the partial derivative 
with respect to $\tau$ ( $x$).  
In order to suppress the finite volume effects we    
work in the {\it continuum formulation}  and   
require, based  on physical considerations,   that the fields satisfy the 
boundary conditions needed 
for the existence of their Fourier transforms 
in the  spatial variable $x^{-}$. 

We note now that  $A_{+}$ appears in the action 
as an {\it auxiliary} field, 
without a kinetic term. It is clear that 
the condensate variable may thus be subtracted out 
from the theory using the frequently adopted 
procedure of  
{\it field redefinition} \cite{nie} on it:  
$\;\;A_{+}\to A_{+}-2\dot\omega/(ae)$, obtaining thereby

\begin{equation}
{\cal L}_{CSM}={\dot\varphi}\varphi'+
{1\over 2}({\dot A}_{-}-{A'_{+}})^{2}+ 2e {\dot\varphi} A_{-}+
a e^{2} A_{+} A_{-},   
\end{equation}
which signals the emergence of a  {\it different structure} of the
Hilbert space compared to that of the SM 
 where we had instead \cite{pre1}
\begin{equation} 
 L= \int\, dx^{-}\;\Bigl[{\dot\varphi}\varphi'+{1\over 2}
({\dot A}_{-}-{A'_{+}})^{2}-({e\over \sqrt {\pi}})
(A_{+}\varphi'-A_{-}{\dot\varphi})\Bigr]+ 
({e\over \sqrt {\pi}}){\dot\omega}{ h(\tau)}  \nonumber 
\end{equation}
  with  $ h(\tau)=\int dx^{-} A_{-}(\tau,x^{-})$, the 
zero mode associated 
with the gauge field $A_{-}$. We recall that the  {\it condensate} or 
$\theta$-vacua in SM  
emerged due to the presence in the theory 
of {\it three} linearly independent operators: the condensate 
$\omega$, its canonically conjugate $h(\tau)$ and $\varphi$ with the vanishing commutator with the other two while the $H^{lf}$ contained in it only 
the field $\varphi$. The Hilbert space could be described in two fashions. 
Selecting  $\varphi$ abd $ h$ as forming the complete set of operators 
led to the chiral vacua while $\varphi$ together with $\omega$ led to 
the description in terms of the condensate or $\theta$-vacua. 
%; the two descriptions being connected through a unitary transformation.
 
%\end{sloppypar}

%\section{LF Hamiltonian Framework}\label{hamilton}
The {\it Lagrange equations in the CSM}  follow to be  

\begin{eqnarray}
    \partial_{+}\partial_{-}\varphi~&=&~-e \partial_{+}A_{-}, \nonumber \\
    \partial_{+}\partial_{+}A_{-}-\partial_{+}\partial_{-}A_{+} ~&=&~
a e^{2} A_{+} +2 e \partial_{+}\varphi, \nonumber \\
\partial_{-}\partial_{-}A_{+}-\partial_{+}\partial_{-}A_{-}
  ~&=&~ a e^{2} A_{-}. 
\end{eqnarray}
and for $ a\neq 1 $ they lead to: 
%\begin{eqnarray}
%  [\Box+{a^2e^2\over(a-1)}]\; A_{-}&=& 0,\nonumber\\ 
% (2-a)~\partial_{+}A_{-}&=& a\partial_{-}A_{+},\nonumber\\ 
% \Box \varphi&=& -2 e \partial_{+}A_{-}, 
%\end{eqnarray}
%which result in 
\begin{eqnarray}
       \Box G(\tau,x)& = & 0 \nonumber\\
\left[ \Box + {e^2a^2\over (a-1) }\right] E(\tau,x) & =& 0, 
\end{eqnarray}  
where $E=(\partial_{+}A_{-}-\partial_{-}A_{+})$ and 
$G= (E-ae\varphi)$. 
Both the massive and massless scalar  excitations 
are present in the theory and  the tachyons
would be absent in the specrtum if   $a>1$; the case  considered 
in this paper. 
 We will confirm in the Hamiltonian framework below 
that the $E$ and $G$ represent, in fact, the two 
independent field operators on the  LF phase space. 
%The structures of the vacua  in the 
%the SM and CSM come out to be quite different; the $\theta$-vacua 
%are absent in the latter. 

The {\it Dirac procedure} \cite{dir1} as applied to  the very simple action of the 
CSM 
is straightforward. The canonical momenta  are 
$    \pi^{+}\approx 0,     \pi^{-}\equiv E=
 \dot{A}_{-}- A'_{+},   \pi_{\varphi}= {\varphi}'+2e A_{-}$ which
result in 
two primary  weak constraints  
$    \pi^{+} \approx 0 $ and 
$ \Omega_1 \equiv (\pi_{\varphi}-\varphi'-2eA_{-})\approx 0$. A 
secondary constraint 
$    \Omega_2 \equiv \partial_{-} E + 
a e^{2} A_{-} \approx 0$ is shown to emerge 
when we require the $\tau$ independence 
(persistency) of 
$\pi^{+}\approx 0$ employing the preliminary  Hamiltonian
\begin{equation}
    H' = {H_c}^{lf} + \int dx~u_{+}\pi^{+}+\int dx~u_{1}\Omega_1 ,
\end{equation}
where $u_{+}$ and $u_{1}$ are the Lagrange multiplier fields and 
${H_c}^{lf}$ is the canonical Hamiltonian  
\begin{equation}
    {H_c}^{lf} = \int\; dx~\left[~
            \frac{1}{2}{E}^{2} + E A_{+}'
-ae^{2} A_{+}A_{-}
               \right]. 
\end{equation}
and we assume initially the  standard   
equal-$\tau$ Poisson  
brackets : 
 $\{E^{\mu}(\tau,x^{-}),A_{\nu}(\tau,
y^{-}) \}=-\delta_{\nu}^{\mu}\delta (x^{-}-y^{-})$,  
$\{\pi_{\varphi}(\tau,x^{-}),\varphi(\tau,y^{-}) \}
=-\delta(x^{-}-y^{-})$ etc.. 
 The persistency requirement for $\Omega_{1} $ results in an equation for 
determining  $u_{1}$. The procedure is repeated with the 
following extended 
Hamiltonian which includes in it also   the secondary constraint  
\begin{equation}
    {H_e}^{lf} = {H_c}^{lf} + \int dx~u_{+}\pi^{+}+\int dx~u_{1}\Omega_1 
+\int dx~u_{2}\Omega_{2}. 
\end{equation}
No more secondary constraints are seen 
to arise; we are left with the persistency conditions which
determine the multiplier fields  $u_{1}$ and $u_{2}$ while $u_{+}$
remains undetermined. We also  
find\footnote{\baselineskip=12pt 
We make the convention that the first variable in an equal-
$\tau$ bracket refers to the longitudinal coordinate $x^{-}\equiv x$
while the second one to $y^{-}\equiv y$ while  $\tau$ is suppressed.} 
 $(C)_{ij}=\{\Omega_{i},\Omega_{j}\}~=~D_{ij}~$ $ (-2
\partial_{x} \delta(x-y))$ where $i,j =1,2$ and $~D_{11}=1,~D_{22}=a
e^2, ~D_{12}=~D_{21}= -e$ and  that $\pi^{+}$ has 
vanishing brackets with $\Omega_{1,2}$. 
The $\pi^{+}\approx 0 $ is     first class weak 
constraint while 
%; it generates local transformations 
%of $A_{+}$ which leave the 
%$H_{e}$ invariant, $\{\pi^{+}, {H_{e}}^{lf}\}= \Omega_{2}\approx 0$. 
 $\Omega_1$ and $\Omega_2 $, which does not depend on  $A_{+}$ or 
$\pi^{+}$,  are second class ones.

We  go over from the Poisson bracket  
to the  Dirac bracket $\{,\}_{D}$ 
 constructed in relation to the pair,    
 $\Omega_1\approx 0$ and   $\Omega_2\approx 0 $ 

\begin{equation} 
\{f(x),g(y)\}_{D}=\{f(x),g(y)\}-\int\int du dv\;\{f(x),\Omega_{i}(u)\}
(C^{-1}(u,v))_{ij}\{\Omega_{j}(v), g(y)\}. 
\end{equation}
Here $C^{-1}$ is the inverse of $C$ and we find 
$(C^{-1}(x,y))_{ij}=B_{ij}$ $K(x,y)$ with  $~B_{11}=a/(a-1)$,$
~B_{22}=1/[(a-1)e^2]$, $~B_{12}=B_{21}$$= 1/[(a-1)e],$ and 
$K(x,y)=-\epsilon(x-y)/4$.  Some of the Dirac brackets are 
$\{\varphi,\varphi\}_{D}= B_{11} ~K(x,y); 
~\{\varphi,E\}_{D} = e B_{11} ~K(x,y); 
~\{E,E\}_{D} = ae^{2} B_{11} ~K(x,y); 
~\{\varphi, A_{-}\}_{D}=-B_{12}~\delta(x-y)/2; 
~\{A_{-},E\}_{D}=B_{11}~\delta (x-y)/2;
~\{A_{-},A_{-}\}_{D}=B_{12}\partial_{x}~\delta(x-y)/2$  
and the only nonvanishing one involving $A_{+}$ or $\pi^{+}$ is 
$\{A_{+},\pi^{+}\}_{D}= \delta(x-y)$.

The eqns. of motion employ now the Dirac brackets and 
inside them, in view of their very construction \cite{dir1}, we may set 
$\Omega_1=0$ and $\Omega_2=0 $  as strong relations. The 
Hamiltonian is therefore  effectively given by $H_{e}$ 
with the terms involving the multipliers $u_{1}$ and $u_{2}$ 
dropped. The multiplier $u_{+}$ is not determined since the constraint 
$\pi^{+}\approx 0$ 
continues to be first class even when the above 
Dirac bracket is employed. 
The variables $\pi_{\varphi}$ and $A_{-}$ are then removed from the 
theory leaving behind $\varphi$, $E$, $A_{+}$, and $\pi^{+}$  as 
the remaining independent variables. 
The canonical Hamiltonian density reduces to 
${\cal H}_{c}^{lf}= E^2/2 +\partial_{-}(A_{+}E)$ while $\dot A_{+}=
\{A_{+}, H_{e}^{lf}\}_{D}=u_{+}$. The surface term in 
the canonical LF Hamiltonian may be ignored if, 
say, $E (=F_{+-})$ vanishes at 
infinity. The variables $\pi^{+}$ and $A_{+}$ are then seen to describe 
a decoupled (from $\varphi$ and $E$) free theory 
and we may hence drop these variables as well. 
The effective LF  Hamiltonian thus takes the simple form 

\begin{equation}
H_{CSM}^{lf} = {1\over {2}} \int dx \; E^{2}, 
\end{equation}
which is to be contrasted with the one found 
in the conventional treatment \cite{bas, abd}. 
 $E$ and $G$ (or $E$ and $\varphi$) are now the independent variables
on the phase space and the Lagrange equations are verified to be
recovered for them, which assures us of the selfconsistency
\cite{dir1}. 
We  stress that in our 
discussion we do {\it not} employ any gauge-fixing.  The same result
for the Hamiltonian could be alternatively  
obtained\footnote{\baselineskip=12pt  A similar discussion is
encountered also in the LF quantized 
Chern-Simons-Higgs system \cite{pre3}.}, 
however,  
if we did introduce  the gauge-fixing constraint $A_{+}\approx 0$ 
and made further 
modification on $\{,\}_{D}$ in order to implement  $A_{+}\approx 0, 
\pi^{+}\approx0$ as well.  
That it is accessible  on the phase space to take care of the 
remaining first class constraint, but not in the bosonized Lagrangian,  
follows from the Hamiltons 
eqns. of motion. We recall \cite{pre1} that in the SM
 $\varphi$, $\omega$, and $\pi_{\omega}=
(e/\sqrt {\pi})\int dx A_{-}$ were shown to be the 
independent operators 
 and that the matter field $\varphi$ appeared instead in the LF 
Hamiltonian.
%\section{Quantization. Vacuum structure in CSM}\label{quant}
\begin{sloppypar}
The  {\it canonical quantization} is peformed 
via the correspondence $i\{f,g\}_{D}\to [f,g] $ and we find the
following equal-$\tau$ commutators 
%\begin{eqnarray} 
%\left[\varphi(x),\varphi(y)\right]& =& i K(x,y)
%{a/(a-1)} ,\nonumber\\ 
%\left[ E(x),E(y) \right]&=& i K(x,y){ a^2 e^2/ (a-1)},\nonumber\\ 
%\left[\varphi(x),E(y)\right]& =& i K(x,y) {ae/ (a-1)} . 
%\end{eqnarray}
%The  algebra gets diagonalized in  terms of 
%$E$ and $G= (E-ea \varphi)$ since 
\begin{eqnarray}
\left[ E(x),E(y) \right]&=& i K(x,y){ a^2 e^2/ (a-1)},\nonumber\\ 
\left[ G(x),E(y)\right]& =& 0,   \nonumber\\
\left[ G(x),G(y)\right]& =& {ia^2 e^2} K(x,y).
\end{eqnarray} 
%\end{sloppypar}
For $a>1$, when the tachyons are absent as seen from (6), 
these commutators  are also 
physical and  the independent field operators $E$ and $G$  generate 
the Hilbert space with a tensor product structure 
of the Fock spaces 
$F_{E}$ and $F_{G}$ of these fields
%, viz.,  ${\cal H}_{E}
%\otimes {\cal H}_{G}$, 
 with  the positive definite metric. 

We can make, in view of (12),  the following LF 
momentum space expansions 
\begin{eqnarray}
E(x,\tau)&=&{ae\over {{\sqrt{(a-1)}}{\sqrt{2\pi}}}} \int_{-\infty}^{\infty} dk
\;{{\theta(k)}\over{\sqrt{2k}}}
\left[d(k,\tau)e^{-ikx}~+~d^{\dag}(k,\tau)e^{ikx}\right],\nonumber\\ 
G(x,\tau) &=& {{ae}\over {\sqrt{2\pi}}} 
\int_{-\infty}^{\infty} dk\;{{\theta(k)}\over{\sqrt{2k}}}
\left[g(k,\tau)e^{-ikx}~+~g^{\dag}(k,\tau)e^{ikx}\right], 
\end{eqnarray}
where the operators ($d, g, d^{\dag},g^{\dag}$) satisfy the 
canonical commutation relations of two independent harmonic
oscillators;   
the well known set of Schwinger's bosonic oscillators, often employed
in the angular momentum theory. The
expression for the Hamiltonian becomes 
\begin{equation}
H_{CSM}^{lf}= \delta(0){{a^2e^2}\over {2(a-1)}}~\int_{k>0}^\infty
{{dk}\over {2k}}
\; N_{d}(k,\tau)
%={{a^2e^2}\over {2(a-1)}}~\int_{k>0}^\infty {{dk}\over {2k}}
%d^{\dag}(k,\tau)d(k,\tau), 
\end{equation}
where we have dropped the infinite zero-point energy term and 
note that \cite{ryd} $\left[d^{\dag}(k,\tau),d(l,\tau)\right]=
-\delta(k-l)$, $d^{\dag}(k,\tau)d(k,\tau)= 
\delta(0) N_{d}(k,\tau)$ etc. 
 with similar expressions 
for the independent g-oscillators. 
We verify that $\left [N_{d}(k,\tau),N_{d}(l,\tau)\right]=0$, 
$\left [N_{d}(k,\tau),N_{g}(l,\tau)\right]=0$,  
$\left [N_{d}(k,\tau),d^{\dag}(k,\tau)\right]=  d^{\dag}(k,\tau) $
etc.. 
\end{sloppypar}

The Fock 
space can hence be built on a basis of eigenstates of the 
hermitian number operators $N_{d}$ and  $N_{g}$. 
The   ground state of CSM is  degenerate and 
described by $\vert 0> = \vert E=0)\otimes \vert G\}$ and it carries 
vanishing LF energy in agreement with the conventional theory discusion 
\cite{bas,abd}.   
For a fixed $k$ these states,  $ \vert E=0)\otimes 
{({g^{\dag}(k,\tau)}^{n}/\sqrt {n!})}\vert 0\}$,  
are labelled by the integers $n=0,1,2,\cdots $. The $\theta$-vacua 
are absent in the CSM. However, we 
recall \cite {pre1} 
that in the SM the degenerate {\it chiral vacua} are also 
labelled  by 
such integers. We remark also that on the LF we work in the Minkowski 
space and that in our discussion we do {\it not} make use of the 
Euclidean space theory action, where the (classical) 
vacuum configurations of
the Euclidean theory gauge field, belonging  to the  
distinct topological sectors, are useful, for example, in the 
functional integral quantization of the gauge theory.   

\newpage
%\bigskip

%\medskip

\nl {\bf Conclusions}\\

\medskip

The LF hyperplane is seen to be {\it equally appropriate} as the conventional
one for quantizing field theory. 
The   {\it front form} 
formulation 
is found to be quite  transparent and the physical contents 
following from the {\it quantized} theory 
agree with those known in the conventional {\it instant form} 
treatment. Evidently, they 
should  not depend on whether we employ the  
conventional or the LF coordinates to span the Minkowski space 
and study the temporal evolution of the quantum dynamical system 
in $t$ or $\tau$ respectively. 

We note that in our context the (LF) 
hyperplanes $x^{\pm}=0$ define the characteristic surfaces of 
hyperbolic partial differential equation.    
It is known from their mathematical theory \cite{sne} 
that a solution exists if we specify the initial data 
on both of the hyperplanes. From the present discussion 
and the earlier works \cite{pre, pre1} we conclude that 
it is sufficient in the {\it front form} treatment to choose 
one of the hyperplanes, as proposed by Dirac \cite{dir},  
for canonically quantizing the theory. 
The equal-$\tau$  {\it commutators of the field operators}, 
 at a fixed 
initial LF-time, form now  a part of the initial data instead and we
deal with operator differential equations. 
The information on the commutators on the other characteristic 
hyperplane seems to be already contained \cite{pre1} 
in the quantized theory and need not be specified separately. 
As a side comment, the well accepted notion 
that a  classical model field 
theory must be upgraded first through quantization, 
before we confront it  with the 
experimental data, finds here a theoretical re-affirmation. 

%We note that in the  {\it quantized} theory 
% needed to solve the 
%operator partial differential equations\footnote{ We recall that in 
%contrast we specify the initial conditions on the field functions in the
%theory of (classical) partial differential equations. 
%In our context  $x^{\pm}=0 $ define the 
%characteristic surfaces of  hyperbolic partial differential 
%equation and as such we would be required 
%to specify the initial data on both the surfaces to obtain 
%solution (e.g., ). 
%The underlying symmetry 
%in $x^{+}, x^{-}$ seems to set strong constraints in the {\it front
%form} quantized theory leading  to appreciable simplifications. 
The {\it physical} Hilbert space is obtained in a direct fashion 
in the LF quantized  CSM and SM gauge theories, once 
the constraints are eliminated and the appreciably 
reduced set of independent operators on the LF phase space
identified. CSM has in it both the massive and the massless 
scalar excitations while only the massive one appears in the 
SM. There are no condensate or $\theta$-vacua in CSM but 
they both have  degenerate vacuum structure. 
In the conventional 
treatment \cite{low} an extended phase space is employed and 
suitable constraints are required to be imposed in order to define the 
{\it physical} Hilbert space which would then lead to the 
description of the physical vacuum state. The functional integral 
method together with the LF quantization may be an efficient tool for
handling the nonperturbative calculations.

A discussion  parallel to 
the one given here can also be made in the {\it front form} theory of
the gauge invariant formulation \cite{abd} of the CSM. 
In an earlier work \cite{pre4}, where  the BRST-BFV functional integral 
quantization was employed, it was demonstrated 
that  this formulation and the gauge noninvariant one  
in fact lead to the same effective action. 
Also the BRS-BFT  quantization method  proposed \cite{bat}
recently can  be extended to the {\it front form} theory as illustrated 
in the Appendix D for the CSM on the LF and where 
different equivalent actions are obtained following the method.

\bigskip

\noindent {\bf Appendix A:  \quad Poincar\'e Generators on the LF} 

\medskip

The Poincar\'e generators in coordinate system 
$\,(x^{0},x^{1},x^{2},x^{3})$,  satisfy 
$[M_{\mu\nu},P_{\sigma}]=-i(P_{\mu}g_{\nu\sigma}-P_{\nu}g_{\mu\sigma})$ and 
$[M_{\mu\nu},M_{\rho\sigma}]=i(M_{\mu\rho}g_{\nu\sigma}+
M_{\nu\sigma}g_{\mu\rho}-M_{\nu\rho}g_{\mu\sigma}-M_{\mu\sigma}g_{\nu\rho})
\,$  where the metric is 
 $g_{\mu\nu}=diag\,(1,-1,-1,-1)$, $\mu=(0,1,2,3)$ and we take 
 $\epsilon_{0123}=\epsilon_{-+12}=1$. If we define 
 $J_{i}=-(1/2)\epsilon_{ikl}M^{kl}$
and  $K_{i}=M_{0i}$, where   $i,j,k,l=1,2,3$, we find 
$[J_{i},F_{j}]=i\epsilon_{ijk}F_{k}\,$ 
for $\,F_{l}=J_{l},P_{l}$ or  $ K_{l}$ while 
$[K_{i},K_{j}]=-i \epsilon_{ijk}J_{k}, \, [K_{i},P_{l}]=-iP_{0}g_{il},
\, [K_{i},P_{0}]=iP_{i},$ and  $[J_{i},P_{0}]=0$.

 The LF  generators are  
$P_{+}$, $P_{-}$, $P_{1} $, $P_{2}$, $M_{12}=-J_{3}$, $M_{+-}=-K_{3}$, $M_{1-}=
-(K_{1}+J_{2})/{\sqrt 2}\,
\equiv {-B_{1}}$, $M_{2-}=-(K_{2}-J_{1})/{\sqrt 2}
\equiv {-B_{2}}$, 
$M_{1+}=-(K_{1}-J_{2})/{\sqrt 2}\equiv -S_{1}$ and 
$M_{2+}=-(K_{2}+J_{1})/{\sqrt 2}\equiv -S_{2}$. We find 
$[B_{1},B_{2}]=0$, $[B_{a},J_{3}]=-i\epsilon_{ab} B_{b}$, 
$[B_{a},K_{3}]=iB_{a}$, $[J_{3},K_{3}]=0$, 
$[S_{1},S_{2}]=0$, $[S_{a},J_{3}]=-i\epsilon_{ab} S_{b}$, 
$[S_{a},K_{3}]=-iS_{a}$ where $a,b=1,2$ and $\epsilon_{12}=-\epsilon_{21}=1$. 
Also $[B_{1},P_{1}]=[B_{2},P_{2}]=i P^{+}$, $[B_{1},P_{2}]=[B_{2},P_{1}]=
0$,  $[B_{a},P^{-}]=iP_{a}$,  $[B_{a},P^{+}]=0$, 
$[S_{1},P_{1}]=[S_{2},P_{2}]=i P^{-}$, 
$[S_{1},P_{2}]=[S_{2},P_{1}]=
0$, $[S_{a},P^{+}]=iP_{a}$,  $[S_{a},P^{-}]=0$,  
$[B_{1},S_{2}]= - [B_{2},S_{2}]=-iJ_{3}$, 
$[B_{1},S_{1}]=[B_{2},S_{2}]=-iK_{3} $. 
For $ P_{\mu}=i\partial_{\mu}$,  and 
$M_{\mu\nu}\to L_{\mu\nu}= i(x_{\mu}\partial_{\nu}-x_{\nu}\partial_{\mu})$ 
we find 
$B_{a}=(x^{+}P^{a}-x^{a}P^{+})$, $S_{a}=(x^{-}P^{a}-x^{a}P^{-})$,  
 $K_{3}=(x^{-}P^{+}-x^{+}P^{-})$  and 
$  \quad J_{3}=(x^{1}P^{2}-x^{2}P^{1})$. 
Under the conventional {\it parity} operation  ${\cal P}$: 
($\;x^{\pm}\leftrightarrow x^{\mp}, x^{1,2}\to 
-x^{1,2}$) and $(  p^{\pm}\leftrightarrow p^{\mp}, p^{1,2}\to 
-p^{1,2}),$ we find  $\vec J\to \vec J,\, \vec K \to -\vec K $, $B_{a}\to -S_{a}$  
etc..
The  six generators 
$\,P_{l}, \, M_{kl}\,$  leave  $x^{0}=0$ hyperplane 
invariant and are  called    
{\it kinematical}  while the remaining $P_{0},\,M_{0k}$ 
the  {\it dynamical } ones. On the LF 
there are {\it seven}  kinematical generators :  $P^{+},P^{1},
P^{2}, B_{1}, B_{2}, J_{3}$ and  $K_{3} $ which leave the LF hyperplane, 
$x^{0}+x^{3}=0$,  invariant and the three {\it dynamical} 
ones $S_{1},S_{2}$ and  $P^{-}$ 
form a mutually commuting set. The $K_{3}$ which was dynamical becomes now 
a kinematical; it generates scale transformations of the LF components of 
$x^{\mu}$, $P^{\mu}$ and $M^{\mu\nu}$.  We note that each of the set 
 $\{B_{1},B_{2},J_{3}\}$ and   $\{S_{1},S_{2},J_{3}\}$ generates 
an  $E_{2}\simeq SO(2)\otimes T_{2} $ algebra; this will be shown below to be 
relevant for 
defining the {\it spin} for massless particle. Including $K_{3}$ in each set 
we find two subalgebras each with four elements. Some useful identities are 
$e^{i\omega K_{3}}\,P^{\pm}\,e^{-i\omega K_{3}}= e^{\pm \omega}\,P^{\pm},
\, e^{i\omega K_{3}}\,P^{\perp}\,e^{-i\omega K_{3}}= P^{\perp}, 
 e^{i\bar v.\bar B}\,P^{-}\,e^{-i\bar v.\bar B}= 
 P^{-}+\bar v.\bar P + {1\over 2}{\bar v}^{2}P^{+},  
 e^{i\bar v.\bar B}\,P^{+}\,e^{-i\bar v.\bar B}= P^{+}, 
 e^{i\bar v.\bar B}\,P^{\perp}\,e^{-i\bar v.\bar B}= 
 P^{\perp}+v^{\perp} P^{+}, 
 e^{i\bar u.\bar S}\,P^{+}\,e^{-i\bar u.\bar S}= 
 P^{+}+\bar u.\bar P + {1\over 2}{\bar u}^{2}P^{-}, 
  e^{i\bar u.\bar S}\,P^{-}\,e^{-i\bar u.\bar S}= P^{-}, 
 e^{i\bar u.\bar S}\,P^{\perp}\,e^{-i\bar u.\bar S}= 
 P^{\perp}+u^{\perp} P^{-}$ 
  where  $P^{\perp}\equiv\bar P=(P^{1},P^{2}), \, v^{\perp}\equiv {\bar v}
= (v_{1},v_{2})\, $ 
and  $(v^{\perp}. P^{\perp})\equiv(\bar v.\bar P)=v_{1}P^{1}+v_{2}P^{2}$ etc. Analogous expressions with 
$P^{\mu}$  replaced by $X^{\mu}$ can be obtained if we use 
$\,[P^{\mu},X_{\nu}]\equiv [i\partial^{\mu},x_{\nu}]= i\delta^{\mu}_{\nu}\,$.

\begin{sloppypar}

\bigskip

\noindent {\bf Appendix B:\qquad  LF Spin Operator. Hadrons in LF Fock
Basis}

\medskip

The Casimir generators of the Poincar\'e group are : $P^{2}\equiv 
P^{\mu}P_{\mu}$ and  $W^{2}$, where  
$W_{\mu}=(-1/2)\epsilon_{\lambda\rho\nu\mu}
M^{\lambda\rho}P^{\nu}$ defines the  Pauli-Lubanski pseudovector. It follows
from $[W_{\mu},W_{\nu}]=i\epsilon_{\mu\nu\lambda\rho} W^{\lambda}P^{\rho}, 
\quad [W_{\mu},P_{\rho}]=0\;$  and  $\;W.P=0$ that in a   
representation charactarized by particular eigenvalues of 
the two Casimir operators we 
may simultaneously diagonalize  $P^{\mu}$ along with just one 
component of  $W^{\mu}$. We have 
$ W^{+} =-[J_{3} P^{+}+B_{1} P^{2}-B_{2} P^{1}], 
W^{-} =J_{3} P^{-}+S_{1} P^{2}-S_{2} P^{1}, 
W^{1} =K_{3} P^{2}+ B_{2} P^{-}- S_{2} P^{+},$ and 
$W^{2} =-[K_{3} P^{1}+ B_{1} P^{-}- S_{1} P^{+}]$ and it shows 
that  $W^{+}$ {\it has a  special place} since it 
contains only the kinematical generators \cite {pre1}. On the LF we define  
  ${\cal J}_{3}= -W^{+}/P^{+}$ as the   {\it spin operator}. 
It may be shown  to 
commute with  $P_{\mu}, B_1,B_2,J_3,$ and  $K_3$. 
For $m\ne 0$ we may use the parametrizations $p^{\mu}:( p^{-}=(m^{2}+
{p^{\perp}}^{2})/(2p^{+}), 
p^{+}=(m/{\sqrt 2})e^{\omega}, 
p^{1}=-v_{1}p^{+},  p^{2}=-v_{2}p^{+})$ and 
${\tilde p}^{\mu}: (1,1,0,0)(m/{\sqrt 2})$ in the rest frame. We have 
$P^{2}(p)= m^{2} I$ and  $W(p)^{2}=
W(\tilde p)^{2}= -m^{2} [J_{1}^2+J_{2}^2+J_{3}^2] = -m^{2} s(s+1) I$ 
where $s$ assumes half-integer values.  
Starting from the rest  state $\vert \tilde p; m,s,\lambda, ..\rangle $
with ${J}_{3}\, \vert\tilde p; m,s,\lambda, ..\rangle
= \lambda \,\vert \tilde p; m,s, \lambda, ..\rangle $ we may build  an 
arbitrary eigenstate of $P^{+}, P^{\perp}, {\cal J}_{3} $  (and $ P^{-}$ ) 
on the LF by 

$$\vert p^{+},p^{\perp}; m,s,\lambda, ..\rangle= e^{i(\bar v. \bar B)} 
e^{-i\omega K_{3}} \vert \tilde p; m,s,\lambda, ..\rangle 
$$ 

\noindent If we make use of the following  {\it identity} \cite{pre}

$${\cal J}_{3}(p)=\;J_{3}+v_{1}B_{2}-v_{2}B_{1}=\quad 
e^{i(\bar v. \bar B)} \;
J_{3} \;e^{-i(\bar v. \bar B)}  $$

\noindent we find ${\cal J}_{3}\, \vert p^{+},p^{\perp}; m,s,\lambda, ..\rangle
= \lambda \,\vert p^{+},p^{\perp};m,s,\lambda, ..\rangle $. 
Introducing  also  the operators ${\cal J}_{a}= 
-({\cal J}_{3} P^{a} + W^{a})/{\sqrt{P^{\mu}P_{\mu}}}$, $a=1,2$, which
do, however, contain 
dynamical generators, we verify that 
$\;[{\cal J}_{i},{\cal J}_{j}]=i\epsilon_{ijk} {\cal J}_{k}$.

For $m=0$ case when $p^{+}\ne0$ 
a convenient parametrization is 
$p^{\mu}:( p^{-}=p^{+} {v^{\perp}}^{2}/2, \,p^{+}, 
p^{1}=-v_{1}p^{+}, p^{2}=-v_{2}p^{+})$ and $\tilde p: 
(0, p^{+}, 0^{\perp})$. We have 
$W^{2}(\tilde p) = -(S_{1}^{2}+S_{2}^{2}){p^{+}}^{2}$ 
and  $[W_{1},W_{2}](\tilde p)=0, \,
[W^{+},W_{1}](\tilde p)=-ip^{+}W_{2}(\tilde p), 
\, [W^{+}, W_{2}](\tilde p)=ip^{+}W_{1}(\tilde p)$ showing that 
$W_{1}, W_{2}$ and  $W^{+}$  generate the algebra 
$SO(2)\otimes T_{2}$. The eigenvalues of 
$W^{2}$ are hence not quantized and they vary continuously. 
This is contrary to the experience so we impose that the physical states 
satisfy in addition  $W_{1,2}
\vert \, \tilde p;\,m=0,..\rangle=0$. Hence 
$ W_{\mu}=-\lambda P_{\mu}$ and  the invariant parameter 
$\lambda $ is taken to define as the {\it spin} of the massless particle.  
From  $-W^{+}(\tilde p)/{\tilde p}^{+}=J_{3}$ we conclude that 
$\lambda$ assumes half-integer values as well.  
We note that   $W^{\mu}W_{\mu}=\lambda^{2} P^{\mu}P_{\mu}=0$  and that 
on the LF the definition of the spin operator  appears  unified 
for massless and massive particles. A parallel discussion based on 
$p^{-}\ne0$ may also be given. 
%the roles of  $W^{+}$ and  $W^{-}$ are  interchanged. 

As an illustration consider the three particle state on the LF with the 
total eigenvalues 
$p^{+}$,  $\lambda$  and $p^{\perp}$. In the {\it standard frame } 
with  $p^{\perp}=0\;$ it may be written as 
($\vert x_{1}p^{+},k^{\perp}_{1}; \lambda_{1}\rangle
\vert x_{2}p^{+},k^{\perp}_{2}; \lambda_{2}\rangle
\vert x_{3}p^{+},k^{\perp}_{3}; \lambda_{3}\rangle$ ) 
with  $\sum_{i=1}^{3} \,x_{i}=1$, $\sum_{i=1}^{3}\,k^{\perp}_{i}=0$, and 
$\lambda=\sum_{i=1}^{3}\,\lambda_{i}$. Aplying 
$e^{-i{(\bar p.\bar B)/p^{+}}}$ on it we obtain 
($\vert x_{1}p^{+},k^{\perp}_{1}+x_{1}p^{\perp}; \lambda_{1}\rangle
\vert x_{2}p^{+},k^{\perp}_{2}+x_{2}p^{\perp}; \lambda_{2}\rangle
\vert x_{3}p^{+},k^{\perp}_{3}+x_{3}p^{\perp}; \lambda_{3}\rangle $ ) 
now with  $p^{\perp}\ne0$. The  $x_{i}$ and $k^{\perp}_{i}$ indicate 
relative (invariant) parameters and do not depend upon the 
reference frame. The  $x_{i}$ is 
the fraction of the total longitudinal momentum 
carried by the  $i^{th} $ particle while 
 $k^{\perp}_{i}$ its transverse momentum. The state of a pion with 
 momentum ($p^{+},p^{\perp}$),  for example, 
 may be expressed  as 
an expansion over the LF Fock states constituted by the different 
number of partons \cite{bro}

$$
\vert \pi : p^{+},p^{\perp} \rangle=
\sum_{n,\lambda}\int {\bar \Pi}_{i}{{dx_{i}d^{2}{k^{\perp}}_{i}}\over
{{\sqrt{x_{i}}\,16\pi^{3}}}} \vert n:\,x_{i}p^{+},x_{i}p^{\perp}+
{k^{\perp}}_{i},
\lambda_{i}\rangle\;\psi_{n/\pi}(x_{1},{k^{\perp}}_{1},
\lambda_{1}; x_{2},...) 
$$

\noindent where the summation is over all the Fock states 
$n$ and spin projections  $\lambda_{i}$, with 
${\bar\Pi}_{i}dx_{i}={\Pi}_{i} dx_{i}\; \delta(\sum x_{i}-1), $ 
and ${\bar\Pi}_{i}d^{2}k^{\perp}_{i}={\Pi}_{i} dk^{\perp}_{i} \;
 \delta^{2}(\sum k^{\perp}_{i})$.  The wave function of the 
parton $\psi_{n/\pi}(x,k^{\perp})$ 
indicates the probability amplitude for finding inside the pion 
the partons in the Fock state $n$ carrying 
the 3-momenta 
$(x_{i}p^{+}, x_{i}p^{\perp}+ k^{\perp}_{i}) $. The 
  Fock state of the pion is also 
        off the energy shell : $\,\sum k^{-}_{i} > p^{-}$. 
%Cada component
%de Fock descreve um sistema de partons livres com uma massa cinem\'atica 
%invariante ${\cal M}^{2}={\sum}_{i}^{n} ({k^{\perp}_{i}}^{2}+{m_{i}}^{2})/
%x_{i}$

The {\it discrete symmetry} transformations may also be defined on the 
LF Fock states \cite{bro, pre1}
For example, under the conventional parity ${\cal P} $ 
the  spin operator  ${\cal J}_{3}$ is not left invariant. 
We may rectify this by defining {\it LF Parity operation} by 
${\cal P}^{lf}=e^{-i\pi J_{1}}{\cal P}$. We find 
then  $B_{1}\to -B_{1}, B_{2}\to B_{2}, P^{\pm}\to P^{\pm}, P^{1}\to 
-P^{1}, P^{2}\to P^{2}$ etc. such that 
${\cal P}^{lf}\vert p^{+},p^{\perp}; m,s,\lambda, ..\rangle
\simeq \vert p^{+},-p^{1}, p^{2}; m,s,\,-\lambda, ..\rangle $. Similar
considerations apply for charge conjugation and  time inversion. For example, 
it is straightforward to construct  the free {\it LF Dirac spinor}  
$\chi(p)= 
[\sqrt{2}p^{+}\Lambda^{+}+(m-\gamma^{a}p^{a})\,\Lambda^{-}]\tilde \chi/
{ {\sqrt{\sqrt {2}p^{+}m}}}$ which is also an eigenstate of 
${\cal J}_{3}$ with eigenvalues 
$\pm 1/2$. Here $\Lambda^{\pm}= \gamma^{0}\gamma^{\pm}/{\sqrt 2}=
\gamma^{\mp}\gamma^{\pm}/2=({\Lambda^{\pm}})^{\dagger}$, 
$ (\Lambda^{\pm})^{2}=\Lambda^{\pm}$,  
and $\chi(\tilde p)\equiv \tilde \chi\,$ with  $\gamma^{0}
\tilde \chi= \tilde \chi$. The conventional (equal-time) 
spinor can also be constructed by the  procedure analogous to that 
followed for the LF spinor and it has  the well known form 
$ \chi_{con}(p)=  (m+\gamma.p)\tilde \chi/
{\sqrt{2m(p^{0}+m)}}$. 
Under the conventional parity operation  ${\cal P}: 
\chi'(p')=c \gamma^{0} \chi(p)$ (since we must require 
$\gamma^{\mu}={L^{\mu}}_{\nu}\,S(L)\gamma^{\nu}{S^{-1}}(L)$, etc.). We find 
$\chi'(p)=c 
[\sqrt{2}p^{-}\Lambda^{-}+(m-\gamma^{a}p^{a})\,\Lambda^{+}]\,\tilde \chi
/{\sqrt{\sqrt {2}p^{-}m}}$. For $p\neq\tilde p$ 
it is not proportional to $\chi(p)$ in contrast to the result in 
the case of the usual spinor where 
$\gamma^{0}\chi_{con}(p^{0},-\vec p)=\chi_{con}(p)$ for  $E>0$ (and  
$\gamma^{0}\eta_{con}(p^{0},-\vec p)=-\eta_{con}(p)$ for  $E<0$). 
However, applying parity operator twice we do show 
$\chi''(p)=c^{2}\chi(p)$ hence leading to the usual result 
$c^{2}=\pm 1$. The LF parity operator over spin $1/2$ Dirac spinor is 
${\cal P}^{lf}= c \,(2J_{1})\,\gamma^{0}$ and the corresponding transform 
of $\chi$ is shown to be an  eigenstate of ${\cal J}_{3}$. 

\bigskip

\noindent {\bf Appendix C: \qquad SSB  Mechanism. Continuum Limit of
Discretized LF Quantized Theory. Nonlocality of LF Hamiltonian}.

\medskip

In order to keep the discussion\footnote{see \cite{pre2} and  
Nuovo Cimento A108 
(1995) 35.} simple we would assume $\omega$ to be a consant background field. 
%The value of   $\omega(=\langle 0\vert \phi\vert 0\rangle)$
%will be seen to characterize the corresponding vacuum  state. 
%The translational  invariance of the ground state requires 
  so that 
${\cal L}={\dot\varphi}{\varphi}^\prime-V(\phi)$.  
Dirac procedure is applied now  to 
construct  Hamiltonian field theory which may be quantized. 
We may avoid using 
distribuitions if we   restrict $x$ to a finite interval from  
 $\,-R/2\,$ to $\,R/2\,$. 
 The 
 {\it  limit to the continuum ($R\to\infty\,$)}, however, 
  must \cite{parisi} be taken  later to remove the 
spurious finite volume effects. 
Expanding $\varphi$ by Fourier series we obtain 
        $\phi(\tau,x) \equiv \omega +\varphi(\tau,x)
= \omega+{1\over\sqrt{R}} {q_{0}(\tau)}+
{1\over\sqrt{R}}\;{{\sum}'_{n\ne 0}}\;\;
{q}_n(\tau)\;e^{-ik_n x}$ 
 where $\,k_n=n(2\pi/R)$, 
$\,n=0,\pm 1,\pm 2, ...\,$ and 
the {\it discretized theory} Lagrangian becomes 
$\;i{{\sum}_{n}}\;k_n \,{q}_{-n}\;{{\dot q}}_{n}-
                                  \int dx \; V(\phi)$. 
The momenta conjugate to ${q}_n$ are 
 ${p}_n=ik_n{q}_{-n}$ 
 and the canonical LF Hamiltonian is found to be  $\int \,
 dx \,V(\omega+\varphi(\tau,x))$. 
The primary constraints$^{}$ are thus 
${p}_0 \approx 0\;$ and $\;{\Phi}_n \equiv 
\;{p}_n-ik_n{q}_{-n}\approx 0\;$ for $\,n\ne 0\,$. 
We follow the {\it standard}  Dirac procedure [5] and find 
three   weak constraints  $\,p_{0}\approx 0$, $\,\beta
\equiv \int dx \,V'(\phi) \approx 0$, and $\,\Phi_{n}\approx 0\,$ 
for $\,n\ne 0\,$ 
on the phase space and they are shown to be  second class. 
We find for $n\ne 0$ and $m\ne 0$: 
$\{{\Phi}_n, {p}_0\}=0$,  
$\{{\Phi}_n,{\Phi}_m\} = -2ik_n \delta_{m+n,0}$, 
$\{{\Phi}_n,\beta\} = \{{p}_n,\beta\} 
= -{(1/ \sqrt{R})}\int dx\;\lbrack \,V''(\phi)-V''([{\omega +
q_{0}]/
\sqrt{R}})\,\rbrack \,e^{-ik_nx}\,
\equiv \,-{{\alpha}_{n}/\sqrt{R}}$, 
$\{{p}_0,{p}_0\} = \{\beta,\beta\,\} = 0$, 
$\{{p}_0,\beta\}=-{(1/\sqrt{R})} \break \int dx\;V''(\phi) 
\equiv -{\alpha/\sqrt{R}}$.  
Implement first the pair of constraints $\;{p}_0\approx 0$, 
$\beta \approx 0$ by modifying the Poisson brackets to 
the star  bracket $\{\}^*$  defined by 
$\{f,g\}^* =\{f,g\} - \lbrack \{f,{p}_0\}\;
\{\beta,g\}- (p_0 \leftrightarrow 
\beta)\rbrack ({\alpha/\sqrt{R}})^{-1}$.
We may then set ${p}_0\,=0$ 
and $\beta \,=0$ as  strong equalities. 
We find  by inspection that the brackets 
$\{\}^*\,$ of the remaining  variables coincide with the standard 
Poisson brackets except for the ones involving $q_{0}$ and 
$\,p_{n}\,$ ($n\ne0$):\quad 
$\,\{q_0,{p}_n\}^*\,=
\{q_0,{\Phi}_n\}^*\,=-({\alpha^{-1}}{\alpha}_n)\,$.  
For example, if 
$V(\phi)=\,({\lambda/4})\,{(\phi^2-{m^2}/\lambda)}^2\;$, 
$\lambda\ge 0, m\ne 0\,$ 
we find  $\;\{q_0,{p}_n\}^*\;
[\{\,3\lambda\,({\omega+q_{0}/\sqrt{R}})^{2}-m^{2}\,\}R\,+
6\lambda(\omega+q_{0}/{\sqrt R})\int\, dx \varphi +
\,3\lambda \,\int\,dx\,\varphi^{2}\,]= 
-\,3\lambda\,[\,2(\omega+q_{0}/{\sqrt R})\,{\sqrt R}
 q_{-n}
 +\int \,dx\,\varphi^{2} 
\,e^{ -ik_{n}x} \,] $.

Implement next the constraints $\,\Phi_{n}\approx 0$ 
with $n\neq 0$. We have $ C_{nm}=\{\Phi_{n},\Phi_{m}\}^* 
\break = -2ik_{n}\delta_{n+m,0}$ 
and its inverse is given by  ${C^{-1}}_{nm} =(1/{2ik_{n}})
\delta_{n+m,0} $. The  Dirac bracket which takes care of all 
the constraints is then given by

$$\{f,g\}_{D}\,= \,\{f,g\}^{*}\,-\,{{\sum}'_{n}}\,{1\over {2ik_{n}}}
\{f,\,\Phi_{n}\}^*\,\{\Phi_{-n},\,g\}^* $$  

\noindent where we may now in addition write $\,p_{n}\,=\,ik_{n}
q_{-n}\,$. It is easily shown that 
$\{q_0,{q}_0\}_{D}\,=0, 
\{q_0,{p}_n\}_{D}\,=\{q_0,\,ik_{n}
q_{-n}\}_{D}\,={1\over 2}\,\{q_0,{p}_n\}^*,
\{q_{n},p_{m}\}_{D}\,={1\over 2}\delta_{nm}$. 

The  limit to the continuum, $R\to\infty$ is taken as usual:  
$\Delta=2 ({\pi/{R}})\to dk$, $k_{n}=n\Delta\to k$, $\sqrt{R} 
q_{-n}\to lim_{R\to\infty}
 \int_{-R/2}^{R/2}{dx}\,\varphi(x)\,e^{ik_{n}x}\equiv\,\int_{-\infty}^
{\infty}\,dx\, \varphi(x)\,e^{ikx} = \sqrt{2\pi}{\tilde\varphi(k)}$ 
for all $\,n $,  
$\sqrt{2\pi}
\varphi(x)  =\int_{-\infty}^{\infty} dk \,\tilde\varphi(k) e^{-ikx}$, 
and  $(q_{0}/{\sqrt{R}})\to 0$. 
From $\{\sqrt{R}\,q_{m},\sqrt{R}\,q_{-n}\}_{D} = 
R\,\delta_{nm}/({2ik_{n}})\,$ following from $\{q_{n},p_{m}\}_{D}$ 
for $n,m\ne 0$  
we derive, on using  $\,R\delta_{nm}\to \int_{-\infty}^{\infty}dx e^{i(k-k')x}=
\,{2\pi\delta(k-k')}$, that 
$\{\tilde\varphi(k),\tilde\varphi(-k')\}_{D}\,=\,\delta
(k-k')/(2ik)\,$ 
where  $ k,k'\,\ne 0$. 
If we use the integral representation of the sgn 
function  the well known LF Dirac bracket 
$\{\varphi(x,\tau),\varphi(y,\tau)\}_{D}=-{1\over 4}\epsilon(x-y)$ is
obtained. 
The expressions of  $\{q_{0},p_{n}\}_{D}$ (or $\{q_{0},\varphi'\}_{D}$) 
show that the DLCQ is harder to work with here\footnote{However, we do 
require it if we  use numerical computations on the computer.}.  
The  continuum limit of the {\it constraint eq.} $\beta=0$ is

$$  
 \omega(\lambda\omega^2-m^2)+lim_{R\to\infty} {1\over R}
 \int_{-R/2}^{R/2} dx \Bigl[ \,(3\lambda\omega^2-m^2)\varphi + 
 \lambda (3\omega\varphi^2+\varphi^3 ) \,
\Bigr]=0$$

\noindent while that for the LF Hamiltonian 

$${ P^{-}\,=\int  dx \,\Bigl [\omega(\lambda\omega^2-m^2)\varphi+
{1\over 2}(3\lambda\omega^2-m^2)\varphi^2+
\lambda\omega\varphi^3+{\lambda\over 4}\varphi^4 
\Bigr ]\,}$$

\noindent These results  follow  immediately 
if we worked directly  in the continuum formulation; we do have 
to handle generalized functions now.  In the LF Hamiltonian theory 
we have an additional new ingredient in the form of the 
{\it constraint equation}.  Elimination of 
$\omega $ using it would lead to a {\it nonlocal LF Hamiltonian} 
corresponding to the 
local one in the equal-time formulation.  
At the tree or classical level the integrals appearing in in 
the constraint eq.  
are convergent and when  $R\to\infty$  it leads to 
$V'(\omega)=0$. In 
equal-time theory this is essentially {\it added} to  it 
as an  external constraint based on physical considerations. In 
the renormalized theory \cite{pre2}  the constraint equation 
 describes 
the high order quantum corrections to the tree level value of the condensate.

The quantization is performed via the correspondence$^{}$  
$i\{f,g\}_{D}\to [f,g]$. Hence 
$\varphi(x,\tau)= {(1/{\sqrt{2\pi}})}\int dk\; 
{\theta(k)}\;
[a(k,\tau)$ $e^{-ikx}+{a^{\dag}}(k,\tau)e^{ikx}]/(\sqrt {2k})$, 
were $a(k,\tau)$ and ${a^{\dag}}(k,\tau)$ 
satisfy the canonical equal-$\tau$ commutation relations, 
$[a(k,\tau),{a(k^\prime,\tau)}^{\dag}]=\delta(k-k^\prime)$ etc.. 
The vacuum state is defined 
by  $\,a(k,\tau){\vert vac\rangle}=0\,$, 
$k> 0$ and the tree level  
description of the {\it SSB} is  given as
follows. The values of $\omega=
\,{\langle\vert \phi\vert\rangle}_{vac}\;$ obtained from  
$V'(\omega)=0$ 
the different   vacua in the theory. 
Distinct  Fock spaces corresponding to different values of $\omega$ 
are built as usual by applying the creation operators on the corresponding
vacuum state.  The $\omega=0$ corresponds to a {\it symmetric phase} 
since the Hamiltonian is then symmetric under 
$\varphi\to -\varphi$. For $\omega\ne0$ this symmetry is 
violated and the  system is 
in a {\it broken or asymmetric phase}.

The extension to $3+1$ dimensions and to global continuous symmetry 
is straight-forward\footnote{ Nuovo Cimento A107 (1994) 549 and ref. 
\cite{pre, pre2}}. 
Consider  real scalar fields 
$\phi_{a} (a=1,2,..N)\,$ which form  an isovector of global 
internal symmetry group 
 $O(N)$. We now  write 
 $\phi_{a}(x,x^{\perp},\tau)
=\omega_{a}+\varphi_{a}(x,x^{\perp},\tau)$ and 
the Lagrangian density is  ${\cal L}=[{\dot\varphi_{a}}{\varphi'_{a}}-
{(1/ 2)}(\partial_i\varphi_{a})(\partial_i\varphi_{a})-V(\phi)]$,  
where $i=1,2$ indicate the transverse space directions. The Taylor series 
expansion of the constraint equations $\beta_{a}=0$ gives a set of coupled
equations $R\,V'_a(\omega)+
\,V''_{ab}(\omega)\int dx \varphi_{b}+\,
V'''_{abc}(\omega)\int dx \varphi_b\varphi_c/2+...=0$. Its discussion at 
the tree level leads to the conventional theory results. 
The  LF symmetry generators are found to be 
$G_{\alpha}(\tau)=
-i\int d^{2} {x^{\perp}} dx \varphi'_{c}(t_{\alpha})_{cd}\varphi_{d}$ 
$=\,\int d^{2}{k^{\perp}}\,dk \, \theta(k)\, 
{a_{c}(k,{k^{\perp}})
^{\dag}} (t_{\alpha})_{cd} a_{d}(k,{k^{\perp}})$ where 
$\alpha,\beta=1,2,..,N(N-1)/2\, $,  are the group indices, 
$t_{\alpha}$ are hermitian and antisymmetric generators of $O(N)$, and 
${a_{c}(k,{k^{\perp}})^{\dag}}$ ($ a_{c}(k,{k^{\perp}})$) is creation (
destruction)  operator  contained in the momentum space expansion 
of $\varphi_{c}$. These  are to be contrasted with the generators 
in the equal-time theory,  
$ Q_{\alpha}(x^{0})=\int d^{3}x \, J^{0}
=-i\int d^{3}x (\partial_{0}\varphi_{a})(t_{\alpha})_{ab}\varphi_{b} 
-i(t_{\alpha}\omega)_{a}\int d^{3}x 
({{d\varphi_{a}}/ dx_{0}})$.  
All  the symmetry generators thus  
annihilate the LF vacuum  and the SSB is  
now seen  in the 
broken symmetry of the quantized theory Hamiltonian. The criteria for 
the counting of the number of Goldstone bosons on the LF is found 
to be the same as in the conventional theory. 
In contrast, the first term 
on the right hand side of $Q_{\alpha}(x^{0})$ does  annihilate  
the conventional theory vacuum  but 
the second term  gives now  non-vanishing contributions  
for some of the (broken) generators. The  symmetry of the
conventional theory vacuum 
is thereby broken  while the quantum Hamiltonian remains invariant.  
The physical content  of SSB in the {\it instant form}
and the {\it front form}, however, is the same 
though achieved  by different descriptions. Alternative proofs  
 on the LF,  
in two dimensions, can be given  
of the Coleman's theorem related to the absence of 
Goldstone bosons; we are unable to implement 
the second class constraints over the phase space.

\end{sloppypar}

\bigskip

\medskip

\noindent{\bf Appendix D:\qquad 
     BRS-BFT Quantization on the LF of the CSM}\footnote 
{\baselineskip=12pt 
Contribuited paper LP-002, 
Session P 17, {\it International Symposium on Lepton-Photon
Interactions- LP'97}, Desy, Hamburg, July 1997 (available as .ps file
on the Desy database). Presented in a talk given at the 
{\it 8th Workshop on Light-Cone
QCD and Nonperturbative Hadronic Physics}, Lutsen, Minnesota, August 1997.}\\

\medskip

Recently,  it was shown \cite{pre1} that the well known {\it condensate} or 
$\theta$-vacua  in the  
SM could be obtained by a straightforward quantization 
of the theory on the  light-front (LF). 
The procedure adopted was  the one proposed 
earlier in connection with the  {\it front form} description 
 of the SSB as described earlier in these lectures.  
The  scalar field of the equivalent bosonized  
SM is 
separated, based on physical considerations, into the {\it dynamical 
bosonic condensate} and the quantum fluctuation fields. The 
Dirac procedure is then followed in order to 
construct the Hamiltonian formulation and the quantized theory. The 
$\theta$-vacua were shown \cite{pre1} to come out naturally 
along with their continuum
normalization. It is then rather important to understand as to how
and why the vacuum structure in the LF quantized CSM should  come out to
be quite  different; as is  known from the rather elaborate studies on CSM 
 in the conventional framework. We could work with the 
standard Dirac method but the recently proposed BFT procedure which 
is elegant and avoids the computation of Dirac brackets. It would 
thus get tested  on the LF as well and it also allows for constructing 
(new) effective Lagrangian theories.

We convert the two second class constraints of the bosonized
CSM with $a>1$ into first class constraints 
according to the BFT formalism. 
We obtain then the  first class Hamiltonian
from the canonical Hamiltonian 
and recover the DB using Poisson brackets in the extended phase space. 
The corresponding first class Lagrangian is then found by 
performing the momentum integrations in the generating functional.

\begin{center}
{\bf (a) Conversion to First Class
Constrained Dynamical System }
\end{center}

The bosonized CSM model (for  $a>1$) is described 
by the action 

\begin{equation}
    S_{CSM} ~=~ \int d^2x~\left[
                          -\frac{1}{4}F_{\mu \nu}F^{\mu \nu}
                          +\frac{1}{2}\partial_{\mu}\phi\partial^{\mu}\phi
                          +eA_{\nu}(\eta^{\mu \nu}
                          -\epsilon^{\mu \nu})\partial_{\mu}\phi
              +\frac{1}{2}ae^{2}A_{\mu}A^{\mu}~\right],
\end{equation}
where 
$a$ is a regularization ambiguity which enters when we calculate 
the fermionic determinant in the fermionic CSM. 
The action in the LF coordinates takes the form 

\begin{equation}
S_{CSM}~=~\int d^2x^{-}\;\left[{1\over 2} (\partial_{+}A_{-}-\partial_{-}A_{+})^{2}
+\partial_{-}\phi\,\partial_{+}\phi +2 e  A_{-}{\partial_{+} \phi}
+a e^{2} A_{+}A_{-}\right], 
\end{equation}

We now make  the {\it separation},   
in  the scalar field (a generalized function) 
:\quad  $\phi(\tau,x^{-})= \omega (\tau)+\varphi(
\tau,x^{-})$. The Lagrangian density then becomes
\begin{equation}
{\cal L}={1\over 2} (\partial_{+}A_{-}-\partial_{-}A_{+})^{2}
+\partial_{-}\varphi\,\partial_{+}\varphi 
+a e^{2} [A_{+}+{2\over
{ae}}(\partial_{+}\varphi+\partial_{+}\omega)] A_{-}, 
\end{equation} 
We note that the dynamical fields are $A_{-}$ and 
$\varphi$ while $A_{+}$ has no kinetic term. On making a 
redefinition of the (auxiliary) field $A_{+}$ we can recast 
the action on the LF  in the  following form 
\begin{equation}
S_{CSM}~=~ \int dx^{-}\;\left[{\dot\varphi}\varphi'
+{1\over 2}({\dot A}_{-}-{A_{+}}')^{2}
- 2e {\dot A_{-}}\varphi
+a e^{2} A_{+}A_{-}\right],
\end{equation}

The canonical momenta are given by
\begin{eqnarray}
    \pi^{+}~&=&~0, \nonumber \\
    \pi^{-}~&=&~ \dot{A}_{-}~-~ A_{+}'-2e\varphi, \nonumber \\
    \pi_{\varphi}~&=&~ {\varphi}'.
\end{eqnarray}
We follow now the Dirac's standard procedure in order to build an
Hamiltonian framework on the LF. The definition of the canonical momenta
leads to two primary constraints 
\begin{eqnarray}
    \pi^{+} \approx 0, \\
 \Omega_1 \equiv (\pi_{\varphi}-\varphi')\approx 0
\end{eqnarray}
and we derive one secondary constraint 
\begin{equation}
    \Omega_2 \equiv \partial_{-} \pi^{-} + +2e \varphi'+
a e^{2} A_{-} \approx 0.
\end{equation}
This one  follows when we require the $\tau$ independence 
(e.g., the persistency) of 
the primary constraint $\pi^{+}$  with respect to the preliminary  Hamiltonian

\begin{equation}
    H' = {H_c}^{l.f.} + \int dx~u_{+}\pi^{+}+\int dx~u_{1}\Omega_1 ,
\end{equation}
where $H_c$ is the canonical Hamiltonian  
\begin{eqnarray}
    {H_c}^{l.f.} &=& \int\!dx~\left[~
            \frac{1}{2}(\pi^{-}+2e\varphi)^2 + (\pi^{-}+2e\varphi) A_{+}'
-ae^{2} A_{+}A_{-}
               \right],
%        &&~~~~~~ \left. - A_{0}\partial_{1}\pi^1
 %              - \frac{1}{2}ae^{2} \{ (A_0)^2 - (A_1)^2 \}
  %             + \frac{1}{2}e^{2}(A_0 - A_1)^2~\right],
\end{eqnarray}
and we employ the standard equal-$\tau$ Poisson brackets. 
The  $u_{+}$ and $u_{1}$ denote  the Lagrange multiplier fields. The 
persistency requirement for $\Omega_{1} $ give conditions to 
determine  $u_{1}$. The 
Hamiltonian is next extended to include also the secondary constraint
\begin{equation}
    {H_e}^{l.f.} = {H_c}^{l.f.} + \int dx~u_{+}\pi^{+}+\int dx~u_{1}\Omega_1 
+\int dx~u_{2}\Omega_{2} 
\end{equation}
and the procedure is now repeated with respect to the extended Hamiltonian. 
For the case $a>1$, 
no more secondary constraints are seen 
to arise and we are left only with the persistency conditions which
determine the multipliers $u_{1}$ and $u_{2}$ while $u_{+}$ is left
undetermined. We also  
find\footnote{We make the convention that the first variable in an equal-
$\tau$ bracket refers to the longitudinal coordinate $x^{-}\equiv x$
while the second one to $y^{-}\equiv y$} 
 $\{\Omega_{i},\Omega_{j}\}~=~D_{ij}~$ $ (-2
\partial_{x} \delta(x-y))$ where $i,j =1,2$ and $~D_{11}=1,~D_{22}=a
e^2, ~D_{12}=~D_{21}= -e$ and  $\pi^{+}$ is shown to have 
vanishing brackets with $\Omega_{1,2}$. 
The $\pi^{+}\approx 0 $ constitutes a  first class constraint on the phase
space; it generates local transformations of $A_{+}$ which leave the 
$H_{e}$ invariant, $\{\pi^{+}, H_{e}\}= \Omega_{2}\approx 0$. 
The  $\Omega_1,\Omega_2 $ constitute a set of 
second class constraints and do not involve $A_{+}$ or $\pi^{+}$.  
It is very convenient, though not necessary, to add to the set of
constraints on the phase space the (accessible) gauge fixing constraint
$A_{+}\approx 0$. It is evident  from  
that such a gauge freedom is {\it  not} 
available at the Lagrangian level. 
%and we will comment on it below in Sec. 3. 
We will also implement  (e.g., turn  into
strong equalities)  the (trivial) 
pair of weak constraints $A_{+}\approx 0,\; \pi^{+}\approx 0$ by defining 
the Dirac brackets with respect to them.  It is  easy to see that 
for the other remaining dynamical variables the corresponding  
Dirac brackets  coincide with the standard Poisson brackets. The
variables $A_{+}, \pi^{+}$ are thus 
removed from the discussion,   leaving 
behind a constrained dynamical system with the two 
second class  constraints $\Omega_{1}, 
\;\Omega_{2}$ and the  light-front Hamiltonian  
\begin{equation}
    H^{l.f.} = \frac{1}{2} \int\!dx~
            (\pi^{-}+2e\varphi)^2 + 
                  \int dx~u_{1}\Omega_1 +\int dx~u_{2}\Omega_{2} 
\end{equation}
which will be now handled by the BFT procedure. 

We introduce the following linear combinations $\top_{i}$, $i=1,2$, 
of the above constraints  

\begin{eqnarray}
    \top_{1}=c_1 (\Omega_1+\frac{1}{M}\Omega_2)\nonumber \\
\top_{2}=c_2 (\Omega_1-\frac{1}{M}\Omega_2)
\end{eqnarray}
where $c_1={1}/{\sqrt{2(1-e/M)}}$,  $c_2={1}/{\sqrt{2(1+e/M)}}$, 
$M^2=a e^2$, and $a>1$. They satisfy 
\begin{equation}
\{\top_i,\top_j\}=\delta_{ij} (-2\partial_{x}\delta(x-y))
\end{equation}
and thus diagonalize the constraint algebra. 

We now introduce new auxiliary fields $\Phi^{i}$
in order to convert the second class constraint $\top_{i}$ into
first class ones in the extended phase space. 
Following BFT \cite{bat} we require these fields to satisfy
\begin{eqnarray}
   \{A^{\mu}(\mbox{or}~ \pi_{\mu}), \Phi^{i} \} &=& 0,~~~
   \{\varphi(\mbox{or}~ \pi_{\varphi}), \Phi^{i} \} = 0, \\ \nonumber
   \{ \Phi^i(x), \Phi^j(y) \} &=& \omega^{ij}(x,y) =
                      -\omega^{ji}(y,x),
\end{eqnarray}
where $\omega^{ij}$ is a constant and antisymmetric matrix. 
The strongly involutive modified constraints 
$\widetilde{\top}_{i}$ satisfying the abelian algebra 
\begin{eqnarray}
\{\widetilde{\top}_{i}, \widetilde{\top}_{j} \}=0
\end{eqnarray}
as well as the boundary conditions,
$\widetilde{\top}_i \mid_{\Phi^i = 0} = \top_i$
are then postulated to take the form of the following expansion 
\begin{equation}
  \widetilde{\top}_i( A^\mu, \pi_\mu, \varphi, \pi_{\varphi}; \Phi^j)
         =  \top_i + \sum_{n=1}^{\infty} \widetilde{\top}_i^{(n)}, 
                       ~~~~~~\top_i^{(n)} \sim (\Phi^j)^n.
\end{equation}
The first order correction terms in this  infinite series  are
written  as 
\begin{equation}
  \widetilde{\top}_i^{(1)}(x) = \int dy X_{ij}(x,y)\Phi^j(y).
\end{equation}
The  first class  
constraint algebra  of $\widetilde{\top}_i$ then leads to 
the following condition: 
\begin{equation}
   \{\top_i,\top_j\} + \{ \widetilde{\top}_i^{(1)},
 \widetilde{\top}_i^{(1)}\}=0  
\end{equation}
or 
\begin{equation}
(-2\partial_x\delta(x-y)) \delta_{ij} +
   \int dw~ dz~
        X_{ik}(x,w) \omega^{kl}(w,z) X_{jl}(y,z)
         = 0.
\end{equation}
There is clearly some arbritrariness  
in the appropriate choice of $\omega^{ij}$ and $X_{ij}$ 
which corresponds to the canonical transformation
in the extended phase space. 
We can take without any loss of generality the simple solutions, 
\begin{eqnarray}
  \omega^{ij}(x,y)
         &=& - \delta^{ij} \epsilon(x-y) \nonumber  \\
  X_{ij}(x,y)
         &=&   \delta_{ij} \partial_{x}\delta ( x- y),
\end{eqnarray}
Their inverses are easily shown to be 
\begin{eqnarray}
  {\omega^{-1}}_{ij}(x,y)
         &=& -\frac{1}{2} \delta_{ij} \partial_x\delta(x-y) \nonumber  \\
  ({X^{-1}})^{ij}(x,y)
         &=&  \frac{1}{2} \delta^{ij} \epsilon( x- y),
\end {eqnarray}

With the above choice,  we find up to the first order
\begin{eqnarray}
\widetilde{ \top}_{i}&=&\top_{i}+\widetilde\top^{(1)}_{i} \\ \nonumber
&=&\top_{i} +\partial \Phi^{i},
\end{eqnarray}
and  a strongly first class  constraint algebra
\begin{equation}
  \{\top_{i}+ \widetilde{\top}^{(1)}_{i},
            \top_{j}+ \widetilde{\top}^{(1)}_{j} \} = 0.
\end{equation}
The higher order correction terms (suppressing the
integration operation ) 
\begin{eqnarray}
\widetilde{\top}^{(n+1)}_{i} =- \frac{1}{n+2} \Phi^{l} 
                  {\omega^{-1}}_{lk} ({X^{-1}})^{kj} 
B_{ji}^{(n)}~~~~~~~(n \geq 1)
\end{eqnarray}
with
\begin{eqnarray}
B^{(n)}_{ji} \equiv \sum^{n}_{m=0} 
       \{ \widetilde{\top}^{(n-m)}_{j}, 
          \widetilde{\top}^{(m)}_{i} \}_{(A, \pi, \varphi, \pi_{\varphi} )}+
         \sum^{n-2}_{m=0} 
          \{ \widetilde{\top}^{(n-m)}_{j}, 
              \widetilde{\top}^{(m+2)}_{i} \}_{(\Phi)}
\end{eqnarray}
automatically vanish as a consequence of the proper choice of $\omega^{ij}$ 
made above. 
The Poisson brackets are to be computed here using the standard canonical 
definition for $A_\mu$ and $\varphi$ as postulated above. 
We have now only  the first class constraints 
in the extended phase space and in view of the proper choice  
only $\widetilde\top_{i}^{(1)}$ contributes in the infinite series above. 
\bigskip

\begin{center}
{\bf (b)-  First Class Hamiltonian  and Dirac Brackets}
\end{center}
\medskip

\begin{sloppypar}

We next introduce modified ("gauge invariant") dynamical variables
$\widetilde{F} \equiv (\widetilde{A}_{\mu},\widetilde{\pi}^{\mu},
\widetilde{\varphi}, \widetilde{\pi}_{\varphi} )$ corresponding to 
$F \equiv (A_{\mu},\pi^{\mu},\varphi, \pi_{\varphi} )$ over the phase space 
by requiring the the following strong involution condition for 
$\widetilde{F}$ with the first class
constraints in our extended phase space, viz, 
\begin{eqnarray}
\{ \widetilde{\top}_{i}, \widetilde{F} \} =0
\end{eqnarray}
with 
\begin{eqnarray}
\widetilde{F}(A_{\mu}, \pi^{\mu}, \varphi, \pi_{\varphi}; \Phi^{j} ) &=&
            F + \sum^{\infty}_{n=1} \widetilde{F}^{ (n)},
            ~~~~~~~ \widetilde{F}^{(n)} \sim (\Phi^{j})^{n}
\end{eqnarray}
and which satisfy the boundary conditions,
$\widetilde{F}\mid_{\Phi^i = 0} = F$.

The first order correction terms are easily shown to be given by 
\begin{equation}
\widetilde{F}^{(1)}(x) =  -\int~ du\, dv\, dz~ \Phi^{j}(u) {{\omega}^{-1}}_{jk}
(u,v) {X^{-1}}^{kl}(v,z)~
 \{ \top_{l}(z), F(x) \}_{(A, \pi, \varphi, \pi_{\varphi})}. 
\end{equation}
We find 
\begin{eqnarray}
\widetilde{A}_{-}^{(1)}~&=&~   
     \frac{1}{2M} \partial(c_1 \Phi^{1}-c_2 \Phi^{2}) \nonumber  \\
\widetilde{\pi}^{-{(1)}}~&=&~ \frac {M}{2} (c_1\Phi^{1}-c_2 \Phi^{2}) 
                                                         \nonumber \\
\widetilde{\varphi}^{(1)} ~&=&~ 
                       - \frac{1}{2} (c_1 \Phi^{1}+c_2 \Phi^{2}),\nonumber \\
\widetilde{\pi}_{\varphi}^{(1)} ~&=&~   \frac{1}{2} \partial
\left[c_1 (1-\frac{2e}{M})\Phi^{1}+c_2  (1+\frac{2e}{M}) \Phi^{2}\right]
\end{eqnarray}
where only the combinations $(c_1 \Phi^{1}\pm c_2 \Phi^{2})$ of the 
auxiliary fields are seen to occur. 
Furthermore, since the  modified variables $\widetilde F= F+\widetilde
 F^{(1)}+...$,   
up to the first order corrections,
are found to be strongly involutive
as a consequence of the proper choice made above, 
the higher order correction terms
\begin{eqnarray}
\widetilde{F}^{(n+1)} &=&
                              -\frac{1}{n+1}
                              \Phi^{j}\omega_{jk} X^{kl} G^{(n)}_{l},
\end{eqnarray}
with
\begin{eqnarray}
G^{(n)}_{l} &=& \sum^{n}_{m=0}
                \{ \top_{i}^{(n-m)}, 
                     \widetilde{F}^{(m)}\}_{(A, \pi, \phi, \pi_{\phi} )}
                +   \sum^{n-2}_{m=0}
                \{ \top_{i}^{(n-m)}, \widetilde{F}^{(m+2)}\}_{(\Phi)}
           +  \{ \top_{i}^{(n+1)}, \widetilde{F}^{(1)} \}_{(\Phi)} 
                  \nonumber \\
\end{eqnarray}
again vanish. In principle we may follow similar procedure for 
any functional of the phase space variables; it may get, however, 
involved.

We make a side remark on the Dirac formulation for dealing with the
systems with second class constraints by using  the Dirac
bracket (DB), rather than extending the phase space. In fact, the
Poisson brackets of the modified (gauge invariant) 
variables $\widetilde F$ in the 
BFT formalism are related \cite {bat} to  the DB,   
 which implement  the constraints $\top_i\approx 0$ in the problem
under discussion,  by the relation
$\{f,g\}_{D}= \{\widetilde f,\widetilde g\}\mid_{\Phi^{i}=0}$. In
view of only the linear first order correction in CSM 
the computation of the right hand side is quite simple.  We list some 
of the Dirac brackets 

\begin{eqnarray}
 \{\pi^{-},\pi^{-}\}_{D}&=&\{\widetilde{\pi^{-}}, 
\widetilde{\pi^{-}}\} |_{\Phi=0}  \nonumber \\
   & = & \{\widetilde{\pi^{-}}^{(1)}, \widetilde{\pi^{-}}^{(1)} \}
         =  \frac{a^2 e^2}{(a-1)} (-\frac{1}{4} \epsilon(x-y)),  
      \nonumber  \\
 \{\varphi,\varphi\}_{D} & = &
\{\widetilde{\varphi}, \widetilde{\varphi}\} |_{\Phi=0} \nonumber \\
  &  = & \{\widetilde{\varphi}^{(1)}, \widetilde{\varphi}^{(1)} \} 
    = \frac{a}{(a-1)} (-\frac{1}{4} \epsilon(x-y))
      \nonumber  \\
 \{\varphi,\pi^{-}\}_{D}&=& \{\widetilde{\varphi}^{(1)}, \widetilde{\pi^{-}}^
{(1)}\} = \frac{ae}{(a-1) } (-\frac{1}{4} \epsilon(x-y)).
\end{eqnarray}
The other ones follow on using the now strong relations 
$\Omega_1=\Omega_2=0$ with respect to $\{,\}_{D}$ and from $H^{l.f}$  
it follows that the LF Hamiltonian 
 reduces effectively to 
\begin{equation}
   H_D^{l.f.}= \frac{1}{2}\int dx  \, (\pi^{-}+2e\varphi)^{2}. 
\end{equation}
The first class LF Hamiltonian $\widetilde H$ which satisfies the
boundary condition $\widetilde H\mid_{\Phi^{i}=0}=H_D^{l.f.}$ and is in
strong involution with the constraints  
$\widetilde \top_{i}\;$,  e.g., 
 $\{\widetilde \top_{i},\widetilde H\}=0 $, may be constructed
following the BT procedure  or simply guessed for the CSM. It is given by 
\begin{equation}
\widetilde H= \frac{1}{2}\int dx(\widetilde \pi^{-}+2e\widetilde\varphi)^2
\end{equation}
which is just the expression in of  $H_D^{l.f.}$ 
with field variables $F$ replaced by the
$\widetilde F$ variables, which already commute with the constraints 
$\widetilde T_{i}$. 
We do also check that $\{\widetilde H,\widetilde H\}=0$ and we may
identify $\widetilde H$ with the BRS Hamiltonian. 
This completes the operatorial 
conversion of the original second class system with the Hamiltonian
$H_{c}$ and constraints $\Omega_{i}$ into the first class one with
the Hamiltonian $\widetilde H$ and (abelian) constraints 
$\widetilde T_{i}$.
\end{sloppypar}
\bigskip
\begin{center}
{\bf (c)- First Class Lagrangian}
\end{center}
\medskip

We consider now the partition function of the model in order to
construct the Lagrangian corresonding to $\widetilde H$ in the
canonical Hamiltonian formulation discussed above. 

We start by representing  each of the auxiliary field $\Phi^{i}$ by a pair
of fields $\pi^{i}, \theta^{i},\; i=1,2\; $ defined by 
\begin{equation}
\Phi^{i}=\frac{1}{2}\pi^{i}-\int du\;\; \epsilon(x-u)\;\theta^{i}(u)
\end{equation}
such that $\pi^{i}, \theta^{i}$ satisfy 
\begin{equation}
\{\pi^{i},\theta^{j}\}=-\delta^{ij}\delta(x-y)\ \  \  etc.,
\end{equation}
e.g., the  (standard Heisenberg type) canonical Poisson brackets. 

 The  phase space partition function is given By the
Faddeev formulae 

\begin{equation}
Z=  \int  {\cal D} A_{-}
          {\cal D} \pi^{-}
          {\cal D} \varphi
          {\cal D} \pi_{\varphi}
          {\cal D} \theta^{1}
          {\cal D} \pi^{1}
          {\cal D} \theta^{2}
          {\cal D} \pi^{2}
               \prod_{i,j = 1}^{2} \delta(\widetilde{\top}_i)
                           \delta(\Gamma_j)
                \mbox{det} \mid \{ \widetilde{\top}_i, \Gamma_j \} \mid
                e^{iS},
\end{equation}
where
\begin{equation}
S  =  \int d^2x \left(
               \pi^{-} {\dot A_{-}} +\pi_{\varphi} {\dot \varphi} 
+ \pi^{1} {\dot \theta^{1}} +\pi^{2} {\dot \theta^{2}} - \widetilde {\cal H}
                \right)\equiv \int d^{2}x \; {\cal L},
\end{equation}
with the Hamiltonian density $\widetilde {\cal H}$ corresponding to the 
Hamiltonian 
$\widetilde H$ which is now expressed in terms
of $(\theta^{i}, \pi_{i})$ rather than in terms of  $\Phi^i$.
The gauge-fixing conditions $\Gamma_i$ are chosen
such  that the determinants occurring in
the functional measure are nonvanishing.
Moreover, $\Gamma_i$ may be taken  to be independent of the momenta
so that they correspond to the  Faddeev-Popov type gauge conditions.

We will now verify in the  {\it unitary gauge}, defined by 
the original  second class constraints:\quad    $\Gamma_i\equiv
\Omega_i=0$, i=1,2  being employed in the partition function, 
 do in fact lead to the original 
Lagrangian. We check that the determinants in the functional
measure are non-vanishing and field independent while  the  product of 
delta functionals reduces to 
\begin{equation}
\delta(\pi_{\varphi}-\varphi')\delta({\pi^{-}}'+2e\varphi'+M^2 A_{-})
\delta({\pi^{1}}'-4\theta^{1})\delta({\pi^{2}}'-4\theta^{2})
\end{equation}
Since $\pi_{\varphi}$ is absent from $\widetilde H$ 
we can perform functional integration over it using 
the first delta functional. 
The second delta functional is exponentiated as usual and we name the
integration variable as $A_{+}$ for convenience. The 
functional integral over $\theta^{1}$ and $\theta^{2}$ are 
easily performed due to the presence of the delta functionals and it 
also reduces $\widetilde {\cal H}$ to  
$(\pi^{-}+2e\varphi)^{2}/2$. The  functional integrations over the
then decoupled variables $\pi^{1}$ and $\pi^{2}$ give rise to constant factors
which are absorbed in the normalization. The partition function in
the unitary gauge thus becomes 
\begin{equation}
Z=  \int  {\cal D} A_{-}
          {\cal D} \pi^{-}
          {\cal D} \varphi
          {\cal D} A_{+}
                          e^{iS},
\end{equation}
with 
\begin{equation}
S  =  \int d^2x \left[
               \pi^{-} {\dot A_{-}} +{\varphi}' {\dot \varphi} 
+({\pi^{-}}'+2e\varphi'+M^{2}A_{-}) A_{+} -\frac{1}{2}(\pi^{-}+2e\varphi)^{2}
                \right],
\end{equation}
Performing the shift $\pi^{-}\rightarrow \pi^{-}-2e\varphi$ and
doing subsequently a Gaussian integral over $\pi^{-}$ we obtain the
original bosonized  Lagrangian with $\omega$ eliminated by the field 
redefinition of $A_{+}$. It is interesting to recall  that 
while constructing the LF Hamiltonian framework we eliminated
the variable $A_{+}$  making use of the gauge freedom on the LF
phase space and  it  gave rise to appreciable simplification. 
However, on going
over to the first class Lagrangian formalism using 
 the partition functional  this  variable reappears as it should, since 
the initial bosonized action is not  gauge invariant due to the 
presence of the mass term for
the gauge field.  Making other acceptable choices for 
gauge-functions  we can arrive at different effective Lagrangians for
the system under consideration. It is interesting to recall that in the 
fermionic Lagrangian the right-handed component of the fermionic field 
describes a free field and only the left-handed one is gauged. 
field while only the left component is gauged. 
It is also clear from our discussion that $\widetilde H$ proposed above 
 is not unique and we could modify it so that it still lead to 
the original Lagrangian  in the unitary gauge. 
The corresponding first class Lagrangian
would produce still other gauge-fixed effective Lagrangians. 
It will be interesting to study the models on the LF 
with more flavours and accompanying 
non-abelian gauge symmetry using the BFT-BFV formalism.

%\vspace{2cm}

\newpage

\renewcommand{\baselinestretch}{1}
\baselineskip=13pt

\begin{center}
 {\bf Acknowledgements}

\end{center}

The author would like to thank the organizing committee of BSCG and 
Prof. Mario Novello for the invitation which offered him the 
valuable opportunity to interact with very active researchers 
in Cosmology and Gravitation. Acknowledgement with thanks are due to  
Werner Israel and Stan Brodsky for constructive discussions and for the 
hospitality offered to him at the SLAC.


\begin{thebibliography}{30}

\bibitem{dir} P.A.M. Dirac, Rev. Mod. Phys.  21 (1949) 392.

\bibitem{pre} P.P. Srivastava, {\sl Lightfront quantization of 
field theory} in {\it Topics in Theoretical Physics}, 
{\sl Festschrift for Paulo Leal Ferreira}, eds., V.C. Aguilera-Navarro et.
al., pgs. 206-217,  IFT-S\~ao Paulo, SP, Brasil (1995); 
hep-th/9610044; {\it Lectures on 
  light-front quantized field theory}:\quad    
{\sl Spontaneous symmetry breaking. 
Phase transition in $(\phi^{4})_{2}$ theory}, 
 Proc. XIV Encontro Nacional de Part\'{\i}culas e Campos, 
Caxamb\'u, MG,   pp. 154-192,   
Sociedade Brasileira de F\'{\i}sica, S\~ao Paulo, SP, Brasil, 1993; 
hep-th/ 9312064;   Nuovo Cimento {\bf A107}
(1994) 549; Nuovo Cimento {\bf A108} (1995) 35. 




\bibitem{ryd} see for example S.S. Schweber, {\it 
Relativistic Quantum Field Theory}, Row, Peterson and Co., New York,
1961;  L.H. Ryder, {\it Quantum Field
Theory}, Cambridge University Press, 2nd Edition, 1996.

\bibitem{col} S. Coleman, Commum. Math. Phys., 31 (1973) 259.


\bibitem{bro} S.J. Brodsky, {\it Light-Cone Quantized QCD and Novel
Hadron Phenomenology}, SLAC-PUB-7645, 1997; 
S.J. Brodsky and H.C. Pauli, {\it Light-Cone Quantization and 
QCD}, Lecture Notes in Physics, vol. 396, eds., H. Mitter et. al., 
Springer-Verlag, Berlin, 1991. 
 

\bibitem{ken} K.G. Wilson, Nucl. Phys. B (proc. Suppl.)  17 (1990). 
R.J. Perry, A. Harindranath, and K.G. Wilson, 
Phys. Rev. Lett.  65 (1990)  2959; K.G. Wilson et. al., Phys. Rev. 
D49 (1994) 6720.


\bibitem{dir1} P.A.M. Dirac, {\it Lectures 
in Quantum Mechanics}, Benjamin, New York, 1964; 
Sudarshan, E.C.G.,  
 Mukunda, N., {\it Classical Dynamics: a modern perspective}, 
Wiley, New York 1974; Hanson, A., Regge T., Teitelboim, C., 
{\it Constrained Hamiltonian Systems},
 Acc. Naz. dei Lincei, Roma 1976.



\bibitem{pre1} P.P. Srivastava,  Mod. Phys. Letts. A13 (1998) 1223; 
{\it Lecture on $\theta$-vacua in the LF quantized Schwinger 
model}, in {\it Geometry, Topology and Physics}, pgs. 260-275, 
Eds,: Apanasov, Bradlow, Rodrigues, and Uhlenbeck, 
Walter de Gruyter \& Co, Berlin, New York, 1997; hep-th/9610149. 




\bibitem{wei} S. Weinberg, Phys. Rev.  150 (1966) 1313. 
\bibitem{kog} J.B. Kogut and D.E. Soper, Phys. Rev.  D1 (1970) 2901.
\bibitem{wit} E. Witten, Commun. Math. Phys. 92 (1984) 455; Nucl.
Phys. B223 (1983) 422. 

\bibitem{sne} See for example, 
I.N. Sneddon, {\it Elements of Partial Differential
Equations}, McGraw-Hill, NY, 1957, pg. 111-115. 



\bibitem{bat} I.A. Batalin and I.V. Tyutin, Int. J. Mod. Phys. 
A6 (1991) 3255. 


\bibitem{pauli} H.C. Pauli and S.J. Brodsky, Phys. Rev. D32 (1985) 1993; 
D32 (1985) 2001.



\bibitem{pre2} P.P. Srivastava, {\it LF quantization and SSB} in 
{\it Hadron Physics 94}, pgs. 253-260, Eds. V. Herscovitz et. al., 
World Scientific, Singapore, 1995; hep-th/9412204,9412205. See also 
{\it AIP Conf. Proc.}, 272 (1993) 2125, Ed. J.R. Sanford, papers 
contribuited to the {\sl Intl.  
Conf. on HEP, Dallas, Texas, August 1992}, the footnote 7 and ref. [2].  

\bibitem{simon} Simon B. and Griffiths R.B., Commun. Math. Phys. 33 
(1973) 145. 

\bibitem{jac} R. Jackiw  and R. Rajaraman,  
             Phys. Rev. Lett.  54 (1985) 1219; 54 (1985) 2060(E). 


\bibitem{low} J. L. Lowenstein and J. Swieca, Ann. Physics 
(N.Y.) 68 (1971) 172. 

\bibitem{pre5} P.P. Srivastava, {\it Light-front 
quantized Chiral Schwinger model and its vacuum structure}, 
SLAC-PUB-8016/98, hep-th/9811225; accepted for publication 
in Phys. Lett. B.   

\bibitem{nie} See for example, 
P. van Nieuwenhuizen, Phys. Rep. 68 (1981) 189.



\bibitem{bas} See for example, D. Boyanovsky, Nucl. Phys. B294 (1987) 223;
 A. Bassetto, L. Griguolo and P. Zanca, Phys. Rev. D50 (1994) 1077; 
 ref. \cite{abd}.



\bibitem{abd} See for example, E. Abdalla, M.C. Abdalla and K. Rothe, {\sl 
Non-Perturbative Methods in Two Dimensional Quantum Field Theory}, 
World Scientific, Singapore, 1991. 


\bibitem{pre3} P.P. Srivastava, Europhys. Lett. 33 (1996) 423;  
{\it LF dynamics of Chern-Simons systems}, ICTP, Trieste preprint 
IC/94/305; hep-th/9412239. 



\bibitem{pre4} P.P. Srivastava,  Phys. Lett. B235 (1990) 287. 





\bibitem{parisi} G. Parisi, {\it Statistical Field Theory}, Addison-
Wesley, 1988. 






%\item{[9]} J. Schwinger, Phys. Rev. 128 (1962) 2425.








%1993 and  2001. See also S.J. Brodsky and G.P. Le Page, in {\it 
%Perturbative Quantum Chromodynamics}, ed. A.H. Mueller, World
%Scientific, Singapore, 1989. 






\end {thebibliography}
\end {document}